\documentclass[draftcls, onecolumn, 12pt]{IEEEtran}
\usepackage{amsfonts}
\usepackage{}
\usepackage{mathrsfs}
\usepackage{amssymb}

\usepackage{cite}

\usepackage{graphicx}

\usepackage{algorithm}

\usepackage{algorithmic}

\usepackage{psfrag}

\usepackage{subfigure}

\usepackage{url}
\usepackage{stfloats}  

\usepackage{amsmath}   
\usepackage{array}

\usepackage{xcolor}
\usepackage{soul}

\newcommand{\picspace}{{\vspace{-0.1 in}}}          

\begin{document}

\title{Energy Efficiency with Proportional Rate Fairness in Multi-Relay OFDM Networks}

\author{Ke~Xiong,~\IEEEmembership{Member,~IEEE}, ~Pingyi~Fan,~\IEEEmembership{Senior Member,~IEEE},~Yang Lu,\\and ~Khaled Ben Letaief,~\IEEEmembership{Fellow,~IEEE}\\


\thanks{K.~Xiong and Y. Lu are with the School of Computer \& Information Technology, Beijing Jiaotong University, Beijing 100044, and  the Department
of Electrical Engineering, Tsinghua University,
Beijing 100084, P. R. China,
e-mail:  kxiong@bjtu.edu.cn, kxiong@tsinghua.edu.cn. P. Y. Fan is with Tsinghua National Laboratory for Information Science and Technology, the Department of Electrical Engineering, Tsinghua University, Beijing, R.P. China, 10008, e-mail:  fpy@tsinghua.edu.cn. K. B. Letaief is with the School of Engineering, Hong Kong University of Science \& Technology (HKUST). e-mail:  eekhaled@ece.ust.hk.
}}

\maketitle

\begin{abstract}
This paper investigates the energy efficiency (EE) in multiple relay aided OFDM system, where decode-and-forward (DF) relay beamforming is employed to help the information transmission. In order to explore the EE performance with user fairness for such a system, we formulate an optimization problem to maximize the EE by jointly considering several factors, the transmission mode selection (DF relay beamforming or direct-link transmission), the helping relay set selection, the subcarrier assignment and the power allocation at the source and relays on subcarriers, under nonlinear proportional rate fairness constraints, where both transmit power consumption and linearly rate-dependent circuit power consumption are taken into account. To solve the non-convex optimization problem, we propose a low-complexity scheme to approximate it.
Simulation results demonstrate its effectiveness.
 We also investigate the effects of the circuit power consumption on system performances and observe that with both the constant and the linearly rate-dependent circuit power consumption, system EE grows with the increment of system average channel-to noise ratio (CNR), but the growth rates show different behaviors. For the constant circuit power consumption, system EE increasing rate is an increasing function of the system average CNR, while for the linearly rate-dependent one, system EE increasing rate is a decreasing function of the system average CNR. This observation is very important which indicates that by deducing the circuit dynamic power consumption per unit data rate, system EE can be greatly enhanced. Besides, we also discuss the effects of the number of users and subcarriers on the system EE performance.
\end{abstract}

\section{Introduction}
\subsection{Background}
Green communication has attracting more and more attention\cite{ict,EH,RF}. It was reported
that the information and communication technology (ICT) industries already accounted for about 3\% of the worldwide
energy consumption and contribute
to about 2\% of the global carbon dioxide emissions\cite{ict}.
To achieve green target for the living environments, one effective way is to harvest renewable energy from the surrounding environment, such as solar, wind, tides and RF signals\cite{EH},\cite{RF}. Another effective way is to design energy-efficient communication systems, in which the information bits per unit of energy is maximized as far as possible \cite{EEsurvey}. Actually, making ICT energy-efficient can bring not only notable positive impacts on environment, but
also long-term profitability to telecommunications operators.
Moreover, energy-efficient communications can also help
people reduce the dependence on fossil fuel and ultimately
reach a sustainable prosperity. Energy-efficient
system design has gradually become a new requirement in future wireless
communication industries \cite{EEsurvey}-\cite{EEsurvey2}.

For attaining energy efficiency (EE)
in wireless systems, the system capacity should be enlarged and meanwhile the system energy consumption should be reduced. In the last decade, much effort has
been made  to increase network throughput, where various
advanced technologies, including
orthogonal frequency-division multiple
(OFDM), multiple-input multiple-output
(MIMO) and cooperative relaying, have been widely investigated for wireless networks to provide high spectral efficiency (SE).
Meantime, numerous resource allocation schemes also have
been presented to meet the quality of service
(QoS) and fairness requirements among different
users.  However,
high SE usually implies large
energy consumption, which sometimes greatly leads to the deviation from green design. Therefore, how to design energy-efficient transmission scheme while meeting the users' fairness requirement become a significant task for next generation wireless networks. Particularly,
EE has been adopted as one of the obligatory evaluation metrics, i.e., EE, SE and cost efficiency, in future 5G systems \cite{5G}.

\subsection{Related Work}
Energy-efficient transmission originated from
energy-constrained networks \cite{Gold}, such as wireless
sensor networks, ad hoc networks, and satellite
communications, where wireless devices are
powered by batteries that are not rechargeable
or hard to recharge, so energy consumption rate
must be taken into account. Recently, since both relay and OFDM have been listed as the promising technologies in broadband wireless systems, such as 3GPP-advanced proposals. Many works investigated the EE  problems in relay networks and OFDM networks, see e.g., \cite{RE1,RE2,RE3,OFDM1,OFDM2,OFDM3,OFDM4,OFDM5,OFDM6}. Nevertheless, less work has been done on the EE design  for the relay-assisted OFDM networks. For example, in \cite{RE1,RE2,RE3}, EE was investigated for relay networks, where however, only single carrier relaying was considered and non OFDM was discussed. In \cite{OFDM1,OFDM2,OFDM3,OFDM4,OFDM5,OFDM6}, EE was maximized for one-hop OFDM systems, where non-relaying was involved.

Different from single-carrier relay systems, where only power and
time resources can be allocated, for a multi-carrier relay
system, frequency resource (i.e., subcarriers) also can be jointly optimized to enhance the EE. Therefore, most recently, some works began to investigate the EE for relay-aided OFDM networks.  In \cite{EEROFDM-3nodes} and \cite{Wiley}, the EE was maximized in three-node OFDM two-way and one-way relay systems, respectively, where amplify-and-forward (AF) was employed.
In \cite{EECROFDM}, EE was studied in cognitive radio network, where only single relay was considered and subcarrier selection and power allocation over the subcarriers were jointly optimized to maximize the EE of the systems. In \cite{EEROFDM1}, decode-and-forward (DF) relay was employed to assist
OFDM uplink transmission from multiple users to a base station, where the information transmission over each subcarrier is only allowed to be forwarded by only one relay.
In \cite{EEROFDM2}, multi-relay aided OFDM system was investigated, where however, it assumed that the BS only transmits information to a single user and no subcarrier assignment among multiple users was involved.

\subsection{Motivations}
In this paper, we investigate the EE in multi-relay assisted multi-user OFDM networks, where a base station (BS) transmits information to multiple users over OFDM channels with the help of multiple DF relay nodes. Our goal is to explore the system optimal EE performance behavior with user fairness constraints. To avoid the interference among the users, each subcarrier is only allowed to be assigned to one user. It is worth emphasizing several differences between our work and existing ones as follows. \textit{Firstly}, different from some existing works, see e.g., \cite{EEROFDM2} and \cite{TVTrelayselect}, where only one relay was selected to help the information transmission on a subcarrier, in our considered system, to enhance the system performance, we allow multiple relays to help the BS forward information to each user on a subcarrier. In order to explore the system performance limit and get a better understand of the system, we assume that the BS and the relay nodes know the full channel-state-information (CSI) and the links and all transmitters have a perfect synchronization\footnote{This can be realized
by a separate low rate feedback channel. According to LTE
specifications \cite{LTE}, CSI can be obtained by various means and
then forwarded to the base station, where the resource allocation
policies are made.}.  Thus, for each around of two-hop relaying transmission, in the first time slot the BS firstly broadcasts the information to all relays and the destination user over the subcarrier, and in the second slot, both the BS and the relays perform a cooperative relay beamforming to transmit information to the destination user. \textit{Secondly}, different from existing works on EE in OFDM systems, see e.g., \cite{Wiley} and \cite{EEROFDM1}, where the EE was maximized under only power constraints or minimal rate constraints with no consideration of multi-user fairness. In our work, EE performance is maximized with the proportional rate
fairness among multiple users by considering a set of nonlinear rate ratio constraints. The reason for us to consider the proportional rate fairness is that the downlink fairness is critical for supporting
various multimedia applications in future mobile communication
systems and the proportional rate constraints can guarantee the instantaneous fairness of multiple users. \emph{Thirdly}, since EE is
defined as the transmitted information amount per unit of energy, and in practical systems, energy is not only consumed by transmitting
information, but also by various circuits for signaling, information processing and buffering, thus when the circuit power consumption is taken into account, the relationship between EE and SE is fundamentally changed. Although many existing works considered the effects of  circuit power consumption on the system EE performance, most of them assumed the circuit power as a constant, see e.g., \cite{OFDM6,EEROFDM2,IET,VTC}. Differently, in our work, we adopt the linear sum-rate dependent circuit power model presented in \cite{CircuitPower}, which is composed of a static part and a sum-rate dependent dynamic part. Such a dynamic circuit power consumption model was adopted in some works for non-relaying OFDM networks, see e.g., \cite{OFDM4} and \cite{VTC-fall}, but has not been considered in multi-relay aided OFDM systems.

Based on above differences, in our considered multi-relay multi-user OFDM system, we shall answer the following fundamental questions.
\begin{itemize}
  \item How to determine the optimal helping relay set for each user over different subcarriers? Considering that the multi-relay assisted DF relaying may not always outperforms the direct link (DL) transmission, we shall also answer when should the DF relay beamforming transmission mode be selected and when should the DL transmission mode should be selected.
  \item How to assign the subcarriers to the users with the joint consideration of relay set determination and transmission mode selection with multi-user fairness?
  \item How to optimally allocate power among the source and relay nodes and how to further allocate power over all subcarriers at each node?
  \item How to jointly optimize the relay set selection, the transmission mode selection, the subcarrier assignment and the power allocation among all nodes over all subcarriers with a low-complex scheme?
  \item How does the linear sum-rate dependent circuit power consumption affect the system EE performance?
\end{itemize}

\subsection{Contributions}
The contributions of this paper can be summarized as follows. \textit{First}, the EE performance in the multi-relay multi-user ODFM network with user fairness is studied. To do so, we formulate an optimization problem, where the helping relay set selection, the transmission mode selection, the subcarrier assignment and the power allocation are jointly optimized under the proportional rate constraints. \textit{Second}, as the problem is non-convex and cannot be directly solved, we first analyze it theoretically and obtain some deterministic results. Based on the obtained results, we design a low-complex scheme which is able to find the approximating optimal solution to the optimization problem. Specifically, for a given user-subcarrier pair, a uniform expression of the achievable information rate for the DF relay bemaforming mode and the DL transmission mode is derived at first, with which the optimal helping relay set and the optimal transmission mode can be determined independently of power allocation. Then, the subcarriers are assigned for all users with the determined helping relaying sets and transmission mode under the proportional rate constraints. Next, the optimal power consumption is derived to achieve the maximum EE under the the proportional rate constraints for the system. \emph{Third}, we present some simulation results to show the effectiveness of our proposed resource allocation scheme. It is shown that our proposed resource allocation is very close to the optimum.
We also discuss the effects of the circuit power consumption on system EE performance and observe that with both the constant and the linearly rate-dependent circuit power consumption the system EE grows with the increment of system average channel-gain-to-noise-ratio (CNR), but the growth rates are very different. For the constant circuit power consumption, system EE increasing rate is a increasing function of average CNR while for the linearly rate-dependent one, system EE increasing rate is a decreasing function of average CNR. This observation is insightful which indicates that by deducing the circuit dynamic power consumption per unit data rate, system EE can be greatly enhanced. Besides, we also discuss the effects of the number of users and subcarriers on the system EE performance.

The rest of this paper is organized as follows. In section II, system model is described. Section III formulates the EE maximization problem and analyzes it. Based on Section III, Section IV presents our proposed resource allocation method and discusses its complexity. Section IV  provides some numerical results and finally we summarize this paper in Section VI.

%

\section{System Model}\label{Sec:SYS}
\begin{figure}
\centering
\includegraphics[width=0.55\textwidth]{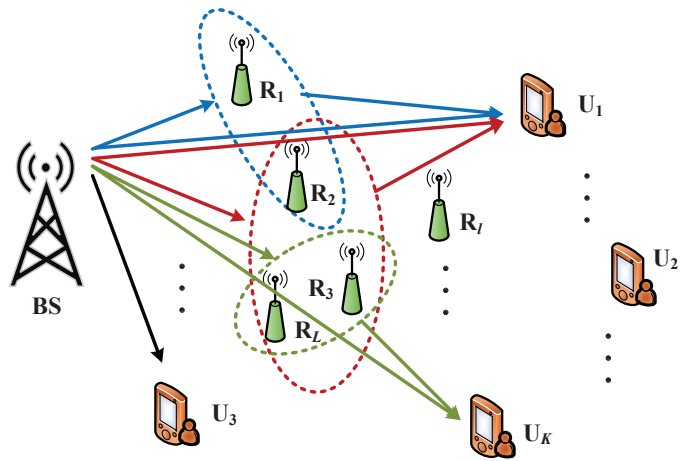}
\caption{System model, where the arrow lines with different colors denote the communications over different subcarrier sets}
\label{Fig:SystemModel}
\end{figure}
Consider a downlink relay-aided multi-user system as shown in Figure \ref{Fig:SystemModel}, where a base station (BS) desires to transmit data to $K$ mobile users. To enhance the system transmission performance, $L$ helping relays are deployed. It is assumed that all nodes in the system are configured
with single omnidirectional antenna.  Half-duplex mode is adopted so that each relay node cannot receive and transmit signals at the same time. To be general, the direct link between the BS and each user is assumed to exist in the system.
With OFDM technology, the total available bandwidth $W$ of the system is divided into $N$ orthogonal subcarriers. Each subcarrier is with the bandwidth of $\frac{W}{N}$ Hz.
For clarity, we use $\textrm{B}$, $n$, $r$ and $k$ to represent the BS, the $n$-th subcarrier, the $r$-th relay and the $k$-th user, respectively, where $1\leq n \leq N$, $1\leq r \leq L$ and $1\leq k \leq K$.

Let $h_{u,v}^{(n)}$ be the channel coefficient of the link between the transmit node $u$ and the receiver node $v$ over subchannel $n$ and ${N_{0}}$ be the noise power spectral density of additive white Gaussian noise, respectively. $\gamma _{u,v}^{(n)}=\frac{|h_{u,v}^{(n)}|^2}{\tfrac{W}{N} N_{0}}$ is the CNR of the subchannel $h_{u,v}^{(n)}$. It is assumed that all subcarriers follow the block fading channel model. That is, the channel coefficient $h_{u,v}^{(n)}$ can be considered as a constant within each time period $T$, but it may vary from one $T$ to the next. To avoid the interference between any two users, a subcarrier is only allowed to be assigned to one user at most within a given $T$. We assume that the subcarrier frequency spacing is wide enough so that the inter subcarrier interference (ICI) can be neglected, which is consistent with the 3GPP LTE standard and related literature, see e.g., \cite{3gpp2}.

\begin{figure}
\centering
\includegraphics[width=0.77\textwidth]{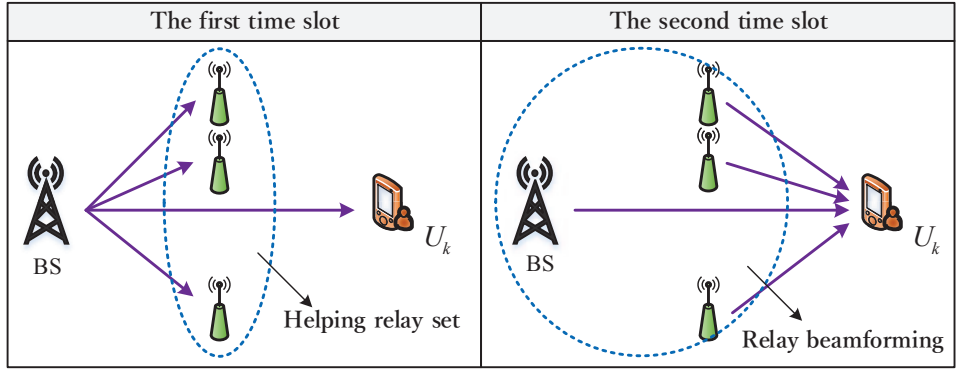}
\caption{Illustration of the multi-relay aided transmission between the BS and user $k$ over subcarrier $n$.}
\label{Fig:TR}
\end{figure}

DF relaying operation is employed at the relay nodes to perform the cooperative relaying transmission. If subcarrier $n$ ($1\leq n \leq N$) is assigned to user $k$ ($1\leq k \leq K$), $k$ is allowed to receive information over $n$ via multiple helping relays.
To implement the DF relaying transmission, each time period $T$ is equally divided into two slots and the transmission protocol can be illustrated in Figure \ref{Fig:TR}.

For each user-subcarrier pair $(k,n)$, in the first slot, the BS broadcasts its information to all relays and user $k$ over subcarrier $n$. The received SNR at relay $r$ and user $k$ over the subchannel $n$ in the first time slot can be, respectively, given by
\begin{equation}\label{Eq:SNRbr}
SNR_{\textrm{B},r}^{(n)} = \gamma _{\textrm{B},r}^{(n)}{P_{\textrm{B}}^{(n)}}
\end{equation}
and
\begin{equation}\label{Eq:SNRbk}
SNR_{\textrm{B},k}^{(n)} = \gamma _{\textrm{B},k}^{(n)}{P_{\textrm{B}}^{(n)}},
\end{equation}
where ${P_{\textrm{B}}^{(n)}}$ is the transmit power at the BS over subcarrier $n$ in the first time slot.

In the second slot, the helping relays decode and then forward the information to user $k$ over subcarrier $n$. Similar to the assumptions in \cite{RBF1,RBF2,RBF3}, the nodes in the network are with block, carrier and symbol synchronization. Therefore, if more than one relay is involved in the relay transmission, they can transmit the decoded information coherently. To enhance the system performance, the BS is also allowed to participate in the transmission of the second slot. Assume that both the channel gain and the channel phase information are known by the transmitters. All involved helping relays and the BS cooperatively can perform a network relay beamforming transmission to maximize the system information transmission performance in the second time slot as shown in Figure \ref{Fig:TR}, which indeed can be regarded as a virtual MISO channel.

Let $\mathcal{R}_{k}^{(n)}$ be the set consisting in the helping relays over subcarrier $n$ for user $k$ in the second slot. Thus, the set composed of all transmit nodes for the MISO channel is given by
$$\mathscr{R}_{k}^{(n)}=\mathcal{R}_{k}^{(n)}\cup \{\textrm{B}\}.$$
As the capacity and optimal
resource allocation over MISO channel is well known in the
literature (e.g., \cite{TSE}), the maximum received SNR at user $k$ over the MISO structure associated with the help of the nodes in $\mathscr{R}_{k}^{(n)}$ can be given by
\begin{equation}\label{Eq:SNRRk}
SNR_{\mathscr{R}_{k}^{(n)}}=\gamma_{\mathscr{R}_{k}^{(n)}}P_{\mathscr{R}_{k}^{(n)}},
\end{equation}
where $\gamma_{\mathscr{R}_{k}^{(n)}}$ is the CNR of the MISO channel which can be calculated by
\begin{equation}
\gamma_{\mathscr{R}_{k}^{(n)}} = \sum\nolimits_{i\in \mathscr{R}_{k}^{(n)}}{\gamma_{i,k}^{(n)}}=\gamma_{\textrm{B},k}^{(n)}+\gamma_{\mathcal{R}_{k}^{(n)}}.
\end{equation}
and $P_{\mathscr{R}_{k}^{(n)}}$ is the available power allocated to MISO structure for the relay beamforming transmission associated with user $k$ on subchannel $n$ in the second slot. Futher, according to \cite{TSE}, to achieve the maximum SNR in (\ref{Eq:SNRRk}), the optimal power allocated to node $i$ in set $\mathscr{R}_{k}^{(n)}$ satisfies
\begin{equation}\label{eq:Prr}
P_{i,k}^{(n)}=\frac{\gamma_{i,k}^{(n)}}{\gamma_{\mathscr{R}_{k}^{(n)}}}P_{\mathscr{R}_{k}^{(n)}},\,\, \textrm{for}\,\,i\in \mathscr{R}_{k}^{(n)}.
\end{equation}

Without loss of generality, we normalize the time period $T$ to be $1$ in the sequel. As a result, the achievable information rate of the DF relaying with the MISO relay beamforming can be given by \cite{DFC}
\begin{flalign}\label{Eq:rknDF}
{r_{k,n}^{(\textrm{DF})}} = \min \Big\{
\tfrac{W}{2N}\log_2 \big(1 + SNR_{\textrm{B},\mathcal{R}_{k}^{(n)}}\big),
\tfrac{W}{2N}\log_2 \big(1 + SNR_{\textrm{B},k}^{(n)} + SNR_{\mathscr{R}_{k}^{(n)}}\big)\Big\},
\end{flalign}
where $\frac{1}{2}$ is the time division factor of the half-duplex relaying, $SNR_{\textrm{B},\mathcal{R}_{k}^{(n)}}$ is the received SNR associated with the relay set $\mathcal{R}_{k}^{(n)}$ and $SNR_{\textrm{B},k}^{(n)}$ can be calculated by (\ref{Eq:SNRbk}).
Moreover, since the achievable information rate with the multi-relay regenerative DF relaying is bounded by the ``de-codable'' information rate in the first slot, thus $SNR_{\textrm{B},\mathcal{R}_{k}^{(n)}}$ is bounded by the minimal SNR over the links between the BS and all relay in set $\mathcal{R}_{k}$, i.e.,
\begin{equation}\label{Eq:SNRbr2}
SNR_{\textrm{B},\mathcal{R}_{k}^{(n)}}=\min\nolimits_{r\in \mathcal{R}_{k}^{(n)}}SNR_{\textrm{B},r}^{(n)} =\min\nolimits_{r\in \mathcal{R}_{k}^{(n)}}\gamma _{\textrm{B},r}^{(n)}{P_{\textrm{B}}^{(n)}}=
\gamma^{(\textrm{min})}_{\textrm{B},\mathcal{R}_{k}^{(n)}}{P_{\textrm{B}}^{(n)}},
\end{equation}
where $\gamma^{(\textrm{min})}_{\textrm{B},\mathcal{R}_{k}^{(n)}}\triangleq \min\nolimits_{r\in \mathcal{R}_{k}^{(n)}}\gamma _{\textrm{B},r}^{(n)}$.

Substituting (\ref{Eq:SNRbr2}), (\ref{Eq:SNRbk}), (\ref{Eq:SNRRk}) into (\ref{Eq:rknDF}), we therefore can express the achievable information rate of user $k$ over the subchannel $n$ with multiple DF helping relays in $\mathcal{R}_{k}^{(n)}$ as
\begin{flalign}\label{Eq:rknDF2}
{r_{k,n}^{(\textrm{DF})}} = \min \Big\{
\tfrac{W}{2N}\log_2 \big(1 + \gamma^{(\textrm{min})}_{\textrm{B},\mathcal{R}_{k}^{(n)}}{P_{\textrm{B}}^{(n)}}\big),
\tfrac{W}{2N}\log_2 \big(1 + \gamma _{\textrm{B},k}^{(n)}{P_{\textrm{B}}^{(n)}} + \gamma_{\mathscr{R}_{k}^{(n)}}P_{\mathscr{R}_{k}^{(n)}}\big)\Big\}.
\end{flalign}

It can be seen that there is a special case, i.e., no helping relay case, in the second time slot. When no relay is involved in the cooperative beamforming, the BS resends the data transmitted in the first time slot to user $k$ over subcarrier $n$ itself. In this case, at the end of the second time slot, user $k$  combines the received signals from the BS in the two time slots by maximum ratio combining (MRC) to perform a better information decoding. This policy can be considered as the repetition-coded direct link transmission. Therefore, the achievable information rate of user $k$ over the subchannel $n$ of the non-relaying direct link transmission is
\begin{flalign}\label{Eq:rknDL}
{r_{k,n}^{(\textrm{DL})}} = \frac{W}{2N}\log_2 \left(1 + 2\gamma _{\textrm{B},r}^{(n)}{P_{\textrm{B}}^{(n)}}\right).
\end{flalign}
The reason we adopt such a repetition-coded transmission policy is to improve the transmission reliability DL link. Another reason is that, by doing so, we can express the achievable information of the DF relay beamforming and the DL mode with a uniformed formulation, which will be explained in Section III.

In the system, for each subcarrier-user pair $(n,k)$, within a given $T$, only one transmission mode of the two modes, i.e., the DF relay beamforming transmission and the DL transmission, is selected to perform the information transmission over subcarrier $n$. Thus, we use a binary variable $\rho_{k,n}\in \{0,1\}$ to indicate that wether the DF relay beamforming mode or the DL mode is selected. Specifically, $\rho_{k,n}=1$ indicates the DF beamforming mode is selected, and $\rho_{k,n}=0$ indicates that the DL mode is selected. Then we can express the end-to-end achievable information rate between the BS and the user $k$ over subcarrier $n$ through the multiple relays as
\begin{flalign}\label{Eq:RnkE2E}
{r_{k,n}^{(\textrm{E2E})}} = (1-\rho_{k,n}){r_{k,n}^{(\textrm{DL})}}+\rho_{k,n}{r_{k,n}^{(\textrm{DF})}},\,\,\rho_{k,n}\in\{0,1\}.
\end{flalign}

As mentioned previously, to avoid the interference among the users, each subcarrier is only allowed to allocate to at most one user. The binary variable $\theta_{k,n}\in \{0,1\}$ is introduced to indicate that wether subcarrier $n$ is assigned to user $k$ or not, where specifically,
$\theta_{k,n}=1$ means that subcarrier $n$ is assigned to user $k$ and $\theta_{k,n}=0$ means that  $n$ is not assigned to user $k$. Thus, 
\begin{equation}\label{Con:theta}
\sum\nolimits_{k=1}^{K}\theta_{k,n}=1,\,\,\forall n,
\end{equation}
and the total end-to-end available information rate for user $k$ over all subcarriers is
\begin{flalign}
R_{k}=\sum\nolimits_{n=1}^{N}{\theta_{k,n}r_{k,n}^{(\textrm{E2E})}}.
\end{flalign}
The total consumed power used for information transmission in the system is the summation of the transmit power at all nodes over all subcarriers, which can be calculated by
\begin{flalign}\label{Eq:Pt}
P_{\textrm{trans}}=\sum\limits_{n=1}^{N}{{P_{\textrm{B}}^{(n)}}+\sum\limits_{n=1}^{N}\sum\nolimits_{k=1}^{K}P_{\mathscr{R}_{k}^{(n)}}}=\sum\limits_{n=1}^{N}{{P_{\textrm{B}}^{(n)}}+\sum\nolimits_{n=1}^{N}\sum\limits_{k=1}^{K}\sum\nolimits_{r\in \mathscr{R}_{k}^{(n)}}{P_{r,k}^{(n)}}}.
\end{flalign}

In the practical communication system, besides the transmit power consumption, the energy consumption also includes circuit
energy consumption caused by signal processing, battery
backup
and active circuit modules (e.g., analog-to-digital converter,
digital-to-analog converter, synthesizer, and mixer) \cite{OFDM4} \cite{VTC-fall}. According to  \cite{OFDM3,OFDM4,VTC-fall}, circuit energy consumption is generally composed of two parts, i.e., the static (fixed) part $P_{\textrm{static}}$ which is used to describe the energy consumed to support the basic circuit operations of the system and the dynamic part which is used for information processing at all nodes and it is dynamically scaled with the sum rate. Let $P_{\textrm{circuit}}$ be the total circuit power comsumption. It can be expressed by
\begin{equation}\label{Eq:pcircuit}
{P_{\textrm{circuit}}} = P_{\textrm{static}} + \xi \sum\nolimits_{k=1}^{K}R_{k},
\end{equation}
where $\xi$ is a constant representing the dynamic power consumption per unit data rate. Thereby, the constant circuit consumption model
used in existing works such as \cite{EEROFDM2,IET,VTC} can be regarded as a special case with $ \xi= 0$ of our adopted dynamic circuit consumption model.

Then the total consumed power of the system can be given by
\begin{equation}\label{Eq:Ptot}
{P_{\textrm{total}}} = \eta P_{\textrm{trans}} + P_{\textrm{circuit}}
\end{equation}
where $\eta$  describes the reciprocal of drain efficiency of the power amplifier \cite{OFDM4} \cite{VTC-fall}, accounting for the power consumption that scales with the transmission (radiated) power due to
amplifier and feeder losses as well as cooling of sites.

The EE (bit/HZ/Joule) of the system then is defined by the ratio of the SE to the consumed total power as
\begin{equation}
EE= \frac{\frac{1}{W}\sum\nolimits_{k=1}^{K}R_{k}}{\eta P_{\textrm{trans}}+P_{\textrm{static}} + \xi \sum\nolimits_{k=1}^{K}R_{k}}.
\end{equation}

\section{Problem Formulation \& Analysis}
\subsection{Problem Formulation}
For the system described in Section \ref{Sec:SYS}, our goal is to investigate its maximum EE performance behavior with multi-user fairness. To this end, we shall jointly optimize several factors, i.e., the transmission mode selection, the multiple relay selection, the subcarrier assignment and the power allocation at the BS and the relays over all subcarriers, with the proportional rate constraints as
\begin{equation}\label{Con:fairness}
{R_1}:{R_2}:...:{R_K} = {\alpha _1}:{\alpha _2}:...:{\alpha _K},
\end{equation}
where ${\alpha _k}$ (k = 1,...,K) is the proportional coefficient, indicating that the data rate among the $K$
users should follow a predetermined proportion. In other
words, the data rate service of the users is performed with a quantized priority.
The benefit of considering such a proportional fairness in the system is that the capacity ratios among users can be explicitly controlled, and each user could be generally ensured to meet its target data rate with sufficient
total available transmit power supply \cite{ProOFDM}. Thus, our EE maximization problem can be mathematically expressed by
\begin{flalign}\label{eq:Opt1}
\max\limits_{\mathcal{R}_{k}^{(n)},\rho_{k,n},\theta_{k,n},{P_{\textrm{B}}^{(n)}},P_{r,k}^{(n)}}&{EE}\\
 \textrm{s.t.}\,\,&(\ref{Con:theta}),(\ref{Con:fairness}),\nonumber\\
 & {P_{\textrm{total}}}\leq P_{\textrm{max}},\nonumber\\
 & {P_{\textrm{B}}^{(n)}}\geq 0, {P_{r,k}^{(n)}}\geq 0,\nonumber
\end{flalign}
where $P_{\textrm{max}}$ represents the maximal available power for the whole system.

It can be seen that the problem in (\ref{eq:Opt1}) is too complex to solve because of the binary variables $\theta_{k,n}$ and $\rho_{k,n}$ and the rate fairness constraints in (\ref{Con:fairness}). Besides, it is difficult to directly observe the properties of the objective function of (\ref{eq:Opt1}). In order to solve the problem (\ref{eq:Opt1}) more efficiently and to better understand the system, we shall first analyze it in the following subsection \ref{subsec:Pa}.

\subsection{Problem Analysis}\label{subsec:Pa}
It is a fact that relaying is usually regarded as an efficient energy saving policy since it
breaks a long distance transmission into several short distance
transmissions. However, this does not mean that relaying always be energy efficient compared with the DL transmission, because the performance behavior of the DF relay transmission closely relies on the channel conditions. As the goal that we employ relay nodes in the transmission is to increase the achievable information rate and the EE performance from the BS to the users, we need to find out when employing the DF relay beamforming is useful. In other words, given the same channel conditions and the same available power, when can the DF relay beamforming support higher data rate than the DL transmission?

Let $P_{k}^{(n)}$ be the available power associated with user $k$ over subcarrier $n$. We have that $P_{k}^{(n)}=P_{\textrm{B}}^{(n)}+\sum \nolimits_{r\in\mathcal{R}_{k}^{(n)}} P_{r,k}^{(n)}$.
With $T=1$, the total available energy $E^{(n)}$ over subcarrier $n$ associated with $T$ satisfies $E^{(n)}=P_{k}^{(n)}$. For DL mode, all transmit energy is consumed by the BS, so ${P_{\textrm{B}}^{(n)}}=P_{k}^{(n)}=E^{(n)}$. For the DF relay beamforming mode, with the same available energy $E^{(n)}$, it satisfies that
\begin{equation}
{E_{\textrm{B}}^{(n)}}+E_{\mathscr{R}_{k}^{(n)}}=E^{(n)},
\end{equation}
where ${E_{\textrm{B}}^{(n)}}$ and $E_{\mathscr{R}_{k}^{(n)}}$ are the assigned energy for the BS in the first time slot and for the nodes of $\mathscr{R}_{k}^{(n)}$ in the second time slot, respectively. As the duration of each time slot is $\tfrac{1}{2}$, we have that
\begin{equation}\label{eq:pebnRn}
{P_{\textrm{B}}^{(n)}}=\tfrac{{E_{\textrm{B}}^{(n)}}}{1/2}=2{E_{\textrm{B}}^{(n)}}\quad
\end{equation}
and
\begin{equation}
P_{\mathscr{R}_{k}^{(n)}}=\tfrac{E_{\mathscr{R}_{k}^{(n)}}}{1/2}=2E_{\mathscr{R}_{k}^{(n)}}=2(E^{(n)}-{E_{\textrm{B}}^{(n)}}).
\end{equation}
for the DF relay beamforming.

Now let us discuss the properties of the system with single relay node and then extend the obtained  properties to multi-relay case.

\subsubsection{Single-relay Case ($L=1$)}For the single-relay system, we obtain the following results.

\textbf{Lemma 1.} \textit{When $L=1$, if the DF relay beamforming  outperforms  the DL mode, it satisfies that $\gamma _{\textrm{B},L}^{(n)}-\gamma _{\textrm{B},k}^{(n)}>0$}.
\begin{proof}
For $L=1$ case, there exists only one relay in the system.
In DL mode, ${r_{k,n}^{(\textrm{DL})}} = \tfrac{W}{2N}\log_2 (1 + 2\gamma _{\textrm{B},k}^{(n)}P_{k}^{(n)})= \tfrac{W}{2N}\log_2 (1 + 2\gamma _{\textrm{B},k}^{(n)}{E^{(n)}})$, while in
DF relay beamforming , according to (\ref{Eq:rknDF}), ${r_{k,n}^{(\textrm{DF})}}$ is bounded by the minimal rate over the two hops. That is, for a given $E^{(n)}$, ${r_{k,n}^{(\textrm{DF})}}\leq\tfrac{W}{2N}\log_2 (1 + \gamma _{\textrm{B},L}^{(n)}{P_{\textrm{B}}^{(n)}})=\tfrac{W}{2N}\log_2 (1 + 2\gamma _{\textrm{B},L}^{(n)}{E_{\textrm{B}}^{(n)}})$
and
${r_{k,n}^{(\textrm{DF})}}\leq \tfrac{W}{2N}\log_2 (1 + \gamma _{\textrm{B},k}^{(n)}{P_{\textrm{B}}^{(n)}}
+\gamma_{\mathscr{R}_{k}^{(n)}}P_{\mathscr{R}_{k}^{(n)}})=\tfrac{W}{2N}\log_2 (1 + 2\gamma _{\textrm{B},k}^{(n)}(P_{k}^{(n)}-E_{\mathscr{R}_{k}^{(n)}})+2\gamma _{\mathscr{R}_{k}^{(n)}}E_{\mathscr{R}_{k}^{(n)}})=\tfrac{W}{2N}\log_2 (1 + 2\gamma _{\textrm{B},k}^{(n)}P_{k}^{(n)}+2(\gamma _{\mathscr{R}_{k}^{(n)}}-\gamma _{\textrm{B},k}^{(n)})E_{\mathscr{R}_{k}^{(n)}})$. When $L=1$, $\gamma _{\mathscr{R}_{k}^{(n)}}=\gamma _{\textrm{B},k}^{(n)}+\gamma _{L,k}^{(n)}$, resulting in that ${r_{k,n}^{(\textrm{DF})}}\leq \tfrac{W}{2N}\log_2 (1 + 2\gamma _{\textrm{B},k}^{(n)}P_{k}^{(n)}+2\gamma _{L,k}^{(n)}E_{\mathscr{R}_{k}^{(n)}})$, which is always larger than $\tfrac{W}{2N}\log_2 (1 + 2\gamma _{\textrm{B},k}^{(n)}P_{k}^{(n)})$
and
$\tfrac{W}{2N}\log_2 (1 + 2\gamma _{\textrm{B},k}^{(n)}E^{(n)})$. Thus, in this case, ${r_{k,n}^{(\textrm{DF})}}$ is only limited by the first term of (\ref{Eq:rknDF}), i.e., ${r_{k,n}^{(\textrm{DF})}} \leq \tfrac{W}{2N}\log_2 (1 + 2\gamma _{\textrm{B},L}^{(n)}{E_{\textrm{B}}^{(n)}})$. It can be deduced that when ${r_{k,n}^{(\textrm{DL})}}<{r_{k,n}^{(\textrm{DF})}}$, it satisfies that $\gamma_{\textrm{B},k}^{(n)}<\gamma_{\textrm{B},L}^{(n)}$. Lemma 1 is thus proved.
\end{proof}

\textbf{Lemma 2.} \textit{When $L=1$, for a given $P_{k}^{(n)}$, the optimal power allocation is}
 \begin{equation}\label{Eq:T2}
\left\{ {\begin{array}{*{20}{l}}
{P_{\textrm{B}}^{(n)}}^*=2P_{k}^{(n)}\left(\tfrac{\gamma_{L,k}^{(n)}+\gamma_{\textrm{B},k}^{(n)}}{\gamma_{\textrm{B},L}^{(n)}+\gamma_{L,k}^{(n)}}\right) \\
{P_{r,k}^{(n)}}^*=2P_{k}^{(n)}\left(\tfrac{\gamma_{\textrm{B},L}^{(n)}-\gamma_{\textrm{B},k}^{(n)}}{\gamma_{\textrm{B},L}^{(n)}+\gamma_{L,k}^{(n)}}\right)\left(\tfrac{\gamma_{r,k}^{(n)}}{\gamma_{\textrm{B},k}^{(n)}+\gamma_{L,k}^{(n)}}\right),\,\, r\in\{\textrm{B},L\}
\end{array}} \right.
 \end{equation}
\textit{where ${P_{\textrm{B}}^{(n)}}^*$  is the optimal power allocated to the BS for the transmission in the first time slot, and ${P_{r,k}^{(n)}}^*$ is the optimal power allocated to the node $r\in \{\textrm{B},L\}$ for the transmission in the second slot,
and the maximum achievable information rate of the DF relay beamforming  is}
\begin{flalign}\label{Eq:EDoptII}
{r_{k,n}^{(\textrm{DF})}}^*=\tfrac{W}{2N}\log_2 \Big(1 +2\beta_{L}^{(n)}  \gamma_{\textrm{B},k}^{(n)}P_{k}^{(n)}\Big),
\end{flalign}
\textit{where}
\begin{equation}
\beta_{L}^{(n)} \triangleq \frac{\gamma_{L,k}^{(n)}\gamma _{\textrm{B},L}^{(n)}+\gamma_{\textrm{B},k}^{(n)}\gamma _{\textrm{B},L}^{(n)}}{\gamma_{\textrm{B},L}^{(n)}\gamma_{\textrm{B},k}^{(n)}+\gamma_{L,k}^{(n)}\gamma_{\textrm{B},k}^{(n)}},
\end{equation}

\begin{proof}For $L=1$ case, in the DF relay beamforming mode, it can be seen that $\tfrac{W}{2N}\log_2 (1 + 2\gamma _{\textrm{B},L}^{(n)}{E_{\textrm{B}}^{(n)}})$ monotonically increases w.r.t ${E_{\textrm{B}}^{(n)}}$ ($0\leq {E_{\textrm{B}}^{(n)}} \leq P_{k}^{(n)}$) while $\tfrac{W}{2N}\log_2 (1 + 2\gamma _{\textrm{B},k}^{(n)}{E_{\textrm{B}}^{(n)}} + (\gamma_{L,k}^{(n)}+\gamma_{\textrm{B},k}^{(n)})(P_{k}^{(n)}-2{E_{\textrm{B}}^{(n)}}))=\tfrac{W}{2N}\log_2 (1 + (\gamma_{L,k}^{(n)}+\gamma_{\textrm{B},k}^{(n)})P_{k}^{(n)}-2\gamma _{L,k}^{(n)}{E_{\textrm{B}}^{(n)}}) $, which monotonically decreases w.r.t ${E_{\textrm{B}}^{(n)}}$. Thus, to maximize ${r_{k,n}^{(\textrm{DF})}}$, the two terms in (\ref{Eq:rknDF}) should equal to each other, i.e., $2\gamma _{\textrm{B},L}^{(n)}{E_{\textrm{B}}^{(n)}}=(\gamma_{L,k}^{(n)}+\gamma_{\textrm{B},k}^{(n)})P_{k}^{(n)}-2((\gamma_{L,k}^{(n)}+\gamma_{\textrm{B},k}^{(n)})-\gamma _{\textrm{B},k}^{(n)}){E_{\textrm{B}}^{(n)}}$. By solving the equation and combining (\ref{eq:pebnRn}),  we  obtain that ${P_{\textrm{B}}^{(n)}}^*=2E_{\textrm{B}}^{(n)}=2P_{k}^{(n)}(\tfrac{\gamma_{L,k}^{(n)}+\gamma_{\textrm{B},k}^{(n)}}{\gamma_{\textrm{B},L}^{(n)}+\gamma_{L,k}^{(n)}})$.
and $P_{\mathscr{R}_{k}^{(n)}}^*=2E_{\mathscr{R}_{k}^{(n)}}=2P_{k}^{(n)}(\tfrac{\gamma_{\textrm{B},L}^{(n)}-\gamma_{\textrm{B},k}^{(n)}}{\gamma_{\textrm{B},L}^{(n)}+\gamma_{L,k}^{(n)}})$.
Further, according to (\ref{eq:Prr}), we can derive the expressions of ${P_{r,k}^{(n)}}^*$ as shown in (\ref{Eq:T2}).
Substituting ${P_{\textrm{B}}^{(n)}}^*$ of (\ref{Eq:T2}) into the first term of (\ref{Eq:rknDF}), we have that
\begin{flalign}\label{Eq:EDoptII}
&{r_{k,n}^{(\textrm{DF})}}^*=\tfrac{W}{2N}\log_2 (1 + \gamma _{\textrm{B},L}^{(n)}{P_{\textrm{B}}^{(n)}}^*)=\tfrac{W}{2N}\log_2 \Big(1 + 2P_{k}^{(n)}\tfrac{(\gamma_{L,k}^{(n)}+\gamma_{\textrm{B},k}^{(n)})\gamma _{\textrm{B},L}^{(n)}}{\gamma_{\textrm{B},L}^{(n)}+\gamma_{L,k}^{(n)}}\Big)\nonumber\\
&=\tfrac{W}{2N}\log_2 \Big(1 +2\gamma_{\textrm{B},k}^{(n)}\tfrac{(\gamma_{L,k}^{(n)}+\gamma_{\textrm{B},k}^{(n)})\gamma _{\textrm{B},L}^{(n)}}{\gamma_{\textrm{B},L}^{(n)}\gamma_{\textrm{B},k}^{(n)}+\gamma_{L,k}^{(n)}\gamma_{\textrm{B},k}^{(n)}}P_{k}^{(n)}\Big)=\tfrac{W}{2N}\log_2 \Big(1 +2\gamma_{\textrm{B},k}^{(n)}\beta_{L}^{(n)}  P_{k}^{(n)}\Big),\nonumber
\end{flalign}
Thus, Lemma 2 is proved.
\end{proof}

As ${P_{\textrm{B}}^{(n)}}\geq 0$ and ${P_{r,k}^{(n)}}\geq 0$, it can be observed from (\ref{Eq:T2}) that it must satisfy that $\gamma_{\textrm{B},L}^{(n)}>\gamma_{\textrm{B},k}^{(n)}$, which is consistent with Lemma 1. Moreover, $\beta_{L}^{(n)}$ can be regarded as the power gain of the DF relay beamforming  transmission between the BS and user $k$ over subcarrier $n$. Therefore, the DF relay beamforming transmission between the BS and user $k$ over subcarrier $n$ with the help of single relay $L$ can  be regarded as a virtual DL link with the achievable information rate of $\tfrac{W}{2N}\log_2 \Big(1 +2\beta_{L}^{(n)}\gamma_{\textrm{B},k}^{(n)} P_{k}^{(n)}\Big)$.

\textbf{Theorem 1.} \textit{When $L=1$, if $\gamma_{\textrm{B},L}^{(n)}-\gamma_{\textrm{B},k}^{(n)}>0$ the DF relay beamforming  transmission mode should be adopted rather than the DL mode in order to achieve a better system performance.}
\begin{proof}
From (\ref{Eq:EDoptII}), it can be seen that for $L=1$ case,  only when $\beta_{L}^{(n)}>1$, the DF relay beamforming  transmission is superior to the DL transmission mode. $\beta_{L}^{(n)}-1=\tfrac{\gamma_{L,k}^{(n)}(\gamma _{\textrm{B},L}^{(n)}-\gamma_{\textrm{B},k}^{(n)})}{\gamma_{\textrm{B},L}^{(n)}\gamma_{\textrm{B},k}^{(n)}+\gamma_{L,k}^{(n)}\gamma_{\textrm{B},k}^{(n)}},
$. Thus, Theorem 1 is proved.
\end{proof}

\subsubsection{Multi-relay Case ($L\geq 2$)}
When $L\geq 2$, all nodes in set $\mathcal{R}_{k}^{(n)}$ can be regarded as a big virtual relay node $\mathcal{R}$ with the effective received SNR from BS to $\mathcal{R}$ being $SNR_{\textrm{B},\mathcal{R}_{k}^{(n)}}=\gamma^{(\textrm{min})}_{\textrm{B},\mathcal{R}_{k}^{(n)}}{P_{\textrm{B}}^{(n)}}$ as shown in (\ref{Eq:SNRbr2}) and with the effective received SNR from $\mathcal{R}$ to user $k$ as shown in (\ref{Eq:SNRRk}).
Similar to the single relay case, for a given $E^{(n)}$,
to maximize ${r_{k,n}^{(\textrm{DF})}}$, the two terms in (\ref{Eq:rknDF}) also should be equal to each other by adjusting the power allocation. Therefore, we can obtain the following optimal power allocation for $L\geq 2$ case in Lemma 3.

\textbf{Lemma 3.} \textit{When $L\geq2$, for a given $P_{k}^{(n)}$, the optimal power allocation is}
\begin{equation}\label{Eq:T3}
\left\{ {\begin{array}{*{20}{l}}
{P_{\textrm{B}}^{(n)}}^*=2P_{k}^{(n)}\left(\tfrac{\gamma_{\mathcal{R}_{k}^{(n)}}+\gamma_{\textrm{B},k}^{(n)}}{\gamma^{(\textrm{min})}_{\textrm{B},\mathcal{R}_{k}^{(n)}}+\gamma_{\mathcal{R}_{k}^{(n)}}}\right)
\\
{P_{r,k}^{(n)}}^*=2P_{k}^{(n)}\left(\frac{\gamma^{(\textrm{min})}_{\textrm{B},\mathcal{R}_{k}^{(n)}}-\gamma_{\textrm{B},k}^{(n)}}{\gamma^{(\textrm{min})}_{\textrm{B},\mathcal{R}_{k}^{(n)}}+\gamma_{\mathcal{R}_{k}^{(n)}}}\right)\frac{\gamma_{r,k}^{(n)}}{\sum\nolimits_{i\in\mathscr{R}_{k}^{(n)}}\gamma_{i,k}^{(n)}},\,\,r\in\mathscr{R}_{k}^{(n)},
\end{array}} \right.
 \end{equation}
\textit{where ${P_{\textrm{B}}^{(n)}}^*$  is the optimal power allocated to the BS for the transmission in the first time slot, and ${P_{r,k}^{(n)}}^*$ is the optimal power allocated to the node $r\in\mathscr{R}_{k}^{(n)}$ for the transmission in the second slot,
and the maximum achievable information rate of the DF relay beamforming  is}
\begin{flalign}\label{Eq:EDoptIIm}
&{r_{k,n}^{(\textrm{DF})^*}}=\tfrac{W}{2N}\log_2 \Big(1 +2\gamma_{\textrm{B},k}^{(n)}\beta_{\mathcal{R}_{k}^{(n)}} P_{k}^{(n)}\Big),\nonumber
\end{flalign}
\textit{where}
\begin{equation}\label{Eq:brkn}
\beta_{\mathcal{R}_{k}^{(n)}}= \frac{\left(\gamma_{\mathcal{R}_{k}^{(n)}}+\gamma_{\textrm{B},k}^{(n)}\right)\gamma^{(\textrm{min})}_{\textrm{B},\mathcal{R}_{k}^{(n)}}}{\gamma^{(\textrm{min})}_{\textrm{B},\mathcal{R}_{k}^{(n)}}\gamma_{\textrm{B},k}^{(n)}+\gamma_{\mathcal{R}_{k}^{(n)}}\gamma_{\textrm{B},k}^{(n)}}.
\end{equation}

\begin{proof}
The proof of Lemma 3 is similar to that of Lemma 2, and Lemma 3 can be obtained by replacing $\gamma_{\textrm{B},L}$, $\gamma_{L,k}$  with $\gamma^{(\textrm{min})}_{\textrm{B},\mathcal{R}_{k}^{(n)}}$ and $\gamma_{\mathcal{R}_{k}^{(n)}}$ of Lemma 2, respectively.
\end{proof}

From Lemma 3, it can be seen that the DF relay beamforming transmission between the BS and user $k$ over subcarrier $n$ for $L\geq 2$ case also can  be regarded as a virtual DL link with the power gain of $\beta_{\mathcal{R}_{k}^{(n)}}$ and the achievable information rate of $\tfrac{W}{2N}\log_2 \Big(1 +2\gamma_{\textrm{B},k}^{(n)}\beta_{\mathcal{R}_{k}^{(n)}} P_{k}^{(n)}\Big)$.

\textbf{Lemma 4.} \textit{When $L\geq2$, if $\gamma^{(\textrm{min})}_{\textrm{B},\mathcal{R}_{k}^{(n)}}>\gamma _{\textrm{B},k}^{(n)}$ the DF relay beamforming  mode is always superior to the DL transmission mode.}
\begin{proof}
It is known that if $\beta_{\mathcal{R}_{k}^{(n)}}-1>0$, the DF relay beamforming over set $\mathcal{R}_{k}^{(n)}$  is superior to the DL mode. As $\beta_{\mathcal{R}_{k}^{(n)}}-1=\tfrac{\gamma_{\mathcal{R}_{k}^{(n)}}\big(\gamma^{(\textrm{min})}_{\textrm{B},\mathcal{R}_{k}^{(n)}}-\gamma_{\textrm{B},k}^{(n)}\big)}{\gamma_{\textrm{B},k}^{(n)}\big(\gamma^{(\textrm{min})}_{\textrm{B},\mathcal{R}_{k}^{(n)}}+\gamma_{\mathcal{R}_{k}^{(n)}}\big)}.$ Therefore, it can be deduced that, as long as $\gamma^{(\textrm{min})}_{\textrm{B},\mathcal{R}_{k}^{(n)}}-\gamma_{\textrm{B},k}^{(n)}>0$, $\beta_{\mathcal{R}_{k}^{(n)}}>1$ holds. Thus, Lemma 4 is proved.
\end{proof}

From Lemma 4, we can easily deduce the following Corollary 1.

\textbf{Corollary 1.} If a helping relay set $\mathcal{R}_{k}^{(n)}$ contains a relay $r$ with $\gamma_{\textrm{B},r}^{(n)}<\gamma_{\textrm{B},k}^{(n)}$, the DF relay beamforming over $\mathcal{R}_{k}^{(n)}$ is inferior to the DL transmission.

\textbf{Theorem 2.} \textit{For a given relay set $\mathcal{R}_{k}^{(n)}=\{1,2,...,r\}$ with $\gamma_{1,k}^{(n)}\geq\gamma_{2,k}^{(n)},...,>\gamma_{r,k}^{(n)}=\gamma^{(\textrm{min})}_{\textrm{B},\mathcal{R}_{k}^{(n)}}$, where $\mathcal{R}_{k}^{(n)}=\{1,2,...,r\}$ and $1\leq r \leq L$, if $\vartriangle\gamma<\Phi_{k,n}$, removing  node $r$  decreases ${r_{k,n}^{(\textrm{DF})^*}}$. Otherwise, if $\vartriangle\gamma>\Phi_{k,n}$, removing node $r$ increases ${r_{k,n}^{(\textrm{DF})^*}}$, where $\vartriangle\gamma=\gamma_{r-1,k}^{(n)}-\gamma_{r,k}^{(n)}$ and}
\begin{flalign}
&\Phi_{k,n}=\tfrac{\left(\gamma_{r,k}^{(n)}\left(\gamma^{(\textrm{min})}_{\textrm{B},\mathcal{R}_{k}^{(n)}}\right)^2 - \gamma_{\textrm{B},k}^{(n)}\gamma_{r,k}^{(n)}\left(\gamma^{(\textrm{min})}_{\textrm{B},\mathcal{R}_{k}^{(n)}}\right)\right)}{\left( \left(\gamma_{\mathcal{R}_{k}^{(n)}}\right)^2 + \gamma_{\textrm{B},k}^{(n)}\gamma_{\mathcal{R}_{k}^{(n)}}-\gamma^{(\textrm{min})}_{\textrm{B},\mathcal{R}_{k}^{(n)}}\gamma_{r,k}^{(n)} - \left(\gamma_{\mathcal{R}_{k}^{(n)}}\right)\gamma_{r,k}^{(n)}\right)}.\nonumber
\end{flalign}

\begin{proof}
Denote the DF relay beamforming power gain associated with $\mathcal{R}_{k}^{(n)}=\{1,2,...,r\}-\{r\}$ as
 $\beta_{\mathcal{R}_{k}^{(n)}}^{-}$.
Therefore,
\begin{flalign}
\beta_{\mathcal{R}_{k}^{(n)}}^{-}-&\beta_{\mathcal{R}_{k}^{(n)}}=\frac{V}{\gamma_{\textrm{B},k}^{(n)}}\frac{1}{\gamma^{(\textrm{min})}_{\textrm{B},\mathcal{R}_{k}^{(n)}}
+\gamma_{\mathcal{R}_{k}^{(n)}}}\cdot\frac{1}{\gamma^{(\textrm{min})}_{\textrm{B},\mathcal{R}_{k}^{(n)}}
+\gamma_{\mathcal{R}_{k}^{(n)}}-\gamma_{r,k}^{(n)}+\vartriangle\gamma},\nonumber
\end{flalign}
where
\begin{flalign}
\Psi=&\left(\gamma_{\mathcal{R}_{k}^{(n)}}\right)^2\vartriangle\gamma - \left(\gamma^{(\textrm{min})}_{\textrm{B},\mathcal{R}_{k}^{(n)}}\right)^2\gamma_{r,k}^{(n)} +\gamma_{\textrm{B},k}^{(n)}\left(\gamma_{\mathcal{R}_{k}^{(n)}}\right)\vartriangle\gamma + \left(\gamma^{(\textrm{min})}_{\textrm{B},\mathcal{R}_{k}^{(n)}}\right)\gamma_{\textrm{B},k}^{(n)}\gamma_{r,k}^{(n)}\nonumber\\ &- \left(\gamma^{(\textrm{min})}_{\textrm{B},\mathcal{R}_{k}^{(n)}}\right)\gamma_{r,k}^{(n)}\vartriangle\gamma - \left(\gamma_{\mathcal{R}_{k}^{(n)}}\right)\gamma_{r,k}^{(n)}\vartriangle\gamma. \end{flalign}
It can be observed that if $\Psi\geq0$, $\beta_{\mathcal{R}_{k}^{(n)}}^{-}\geq \beta_{\mathcal{R}_{k}^{(n)}}$. Otherwise, $\beta_{\mathcal{R}_{k}^{(n)}}^{-}<\beta_{\mathcal{R}_{k}^{(n)}}$. That is, if $\vartriangle\gamma<\Phi_{k,n}$, $\beta_{\mathcal{R}_{k}^{(n)}}^{-}\geq \beta_{\mathcal{R}_{k}^{(n)}}$. Otherwise, $\beta_{\mathcal{R}_{k}^{(n)}}^{-}<\beta_{\mathcal{R}_{k}^{(n)}}$.
\end{proof}

\textbf{Theorem 3.} \textit{For a given relay set $\mathcal{R}_{k}^{(n)}=\{1,2,...,r\}$ with $\gamma_{1,k}^{(n)}\geq\gamma_{2,k}^{(n)},...,>\gamma_{r,k}^{(n)}=\gamma^{(\textrm{min})}_{\textrm{B},\mathcal{R}_{k}^{(n)}}$, where $\mathcal{R}_{k}^{(n)}=\{1,2,...,r\}$, removing node $j\in \{1,2,...,r-1\}$ decreases ${r_{k,n}^{(\textrm{DF})^*}}$.}

\begin{proof}
Denote the DF relay beamforming power gain associated with $\mathcal{R}_{k}^{(n)}=\{1,2,...,r\}-\{j\}$ as
$\beta_{\mathcal{R'}_{k}^{(n)}}^{-}$.
Therefore,
\begin{flalign}
\beta_{\mathscr{R'}_{k}^{(n)}}^{-}-&\beta_{\mathcal{R}_{k}^{(n)}}
=\frac{\gamma^{(\textrm{min})}_{\textrm{B},\mathcal{R}_{k}^{(n)}}}{\gamma_{\textrm{B},k}^{(n)}}\cdot
\frac{\gamma_{j,k}^{(n)}}{\gamma_{\mathcal{R}_{k}^{(n)}}+\gamma^{(\textrm{min})}_{\textrm{B},\mathcal{R}_{k}^{(n)}}}\cdot
\frac{(\gamma_{\textrm{B},k}^{(n)}-\gamma^{(\textrm{min})}_{\textrm{B},\mathcal{R}_{k}^{(n)}})}{\gamma_{\mathcal{R}_{k}^{(n)}}+\gamma^{(\textrm{min})}_{\textrm{B},\mathcal{R}_{k}^{(n)}}-\gamma_{j,k}^{(n)}}
\end{flalign}
Since $(\gamma_{\textrm{B},k}^{(n)}-\gamma^{(\textrm{min})}_{\textrm{B},\mathcal{R}_{k}^{(n)}})<0$ and $\gamma_{\mathcal{R}_{k}^{(n)}}+\gamma^{(\textrm{min})}_{\textrm{B},\mathcal{R}_{k}^{(n)}}-\gamma_{j,k}^{(n)}>0$,
it can be observed that
$\beta_{\mathscr{R'}_{k}^{(n)}}^{-}-\beta_{\mathcal{R}_{k}^{(n)}}<0$ always hold. Therefore, Theorem 3 is proved.
\end{proof}

With the results obtained above, we shall design an efficient algorithm to solve Problem (\ref{eq:Opt1}) in the next Section.

\section{Resource Allocation Scheme Design and Complexity Analysis}

In this section, we shall present our proposed resource allocation scheme for Problem (\ref{eq:Opt1}) at first in subsection \ref{Sec:RA} and then discuss its computational complexity in subsection \ref{Sec:CA}.
\begin{figure}
\centering
\includegraphics[width=0.45\textwidth]{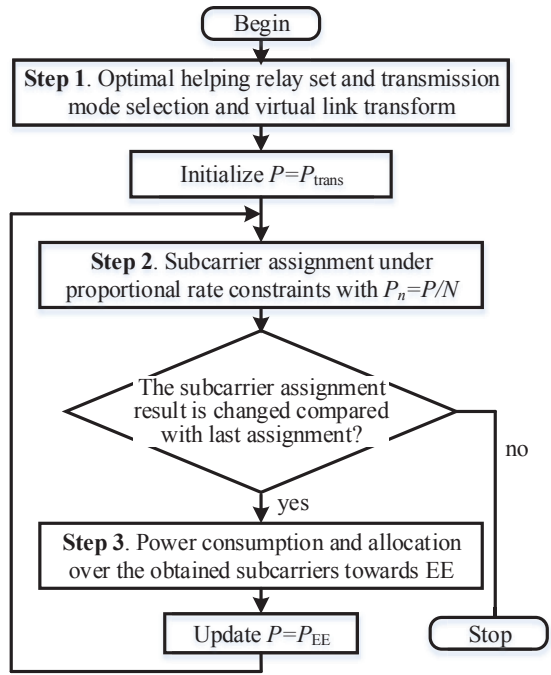}
\caption{Framework of our proposed resource allocation scheme}
\label{Fig:fowchart}
\end{figure}

\subsection{Our Proposed Resource Allocation Scheme}\label{Sec:RA}
The framework of our proposed resource allocation scheme is shown in Figure \ref{Fig:fowchart}, which contains three main steps. In Step 1, it determines the optimal set of helping relay nodes ${\mathcal{R}_{k}^{(n)}}^*$ for each user-subcarrier pair $(k,n)$ and then convert all DF relay beamforming links into virtual DL links. In Step 2, it assigns the subcarriers to the users with an equal power allocation policy undter the rate fairness constraint on the basis of the virtual links obtained in step 1. In Step 3, it finds the optimal power consumption to achieve the maximum EE under the available power constraint $P_{\textrm{max}}$ and then optimally allocate the power to the BS and the relay node in terms of Lemma 3.

\subsubsection{Optimal ${\mathcal{R}_{k}^{(n)}}^*$ and $\rho_{k,n}^*$ and Virtual Link Transformation}
Lemma 2 and Lemma 3 show a very interesting result that for a given user-subcarrier pair $(k,n)$. That is, we can convert the relay-assisted OFDM system into a broadcast channel with  virtual DL links. The difference between the real DL links and the virtual DL links is that the allocated power to the virtual link $P_{k}^{(n)}$ should be further allocated to ${P_{\textrm{B}}^{(n)}}$ and $P_{\mathcal{R}_{k}^{(n)}}$ for the transmissions in the first time slot and the second time slot, respectively. The benefit brought by such a virtual link transformation is that, by doing so, the problem (\ref{eq:Opt1}) can be simplified and better understood since the power could be allocated to the virtual links at first and then to  the BS and the relays for each $n$ with separated operations.

To achieve the optimum EE performance of the system, for given available power, the SE should be maximized. Therefore, it is required to find the maximum power gains of the virtual links for all $(k,n)$ pairs. That is, for each $(k,n)$ pair, a set of relays are required to be selected among $L$ candidate relays to achieve the optimal performance. From the analysis in Section III, we can observe the following facts.
First, according to Corollary 1, the relays whose CNR from the BS is less than $\gamma_{\textrm{B},k}^{(n)}$ should not be involved in the DF relay beamforming transmission. Second, according Theorem 2 and Theorem 3,  for a given relay set $\mathcal{R}_{k}^{(n)}$, if one removes the relay node with the minimal $\gamma_{\textrm{B},r}^{(n)}$ in it, the performance of the DF relay beamforming may be increased, but if we remove the relay node with non-minimal $\gamma_{\textrm{B},r}^{(n)}$ in $\mathcal{R}_{k}^{(n)}$, the performance of the DF relay beamforming will definitely be decreased. To find the optimal ${\mathcal{R}_{k}^{(n)}}^*$, it is only necessary to check the set which is possible to increase $r_{\textrm{B},k}^{(\textrm{E2E})}$. Thus, we only need to compare the set of $\mathcal{R}_{k}^{(n)}$ and $\mathcal{R}_{k}^{(n)}-\{r\}$ where $r$ is the relay with the minimal CNR of the first hop among all relays. Following these observations, we propose the Algorithm \ref{Alg:relay} to find the optimal ${\mathcal{R}_{k}^{(n)}}^*$ for a given $(k,n)$ pair. In Algorithm, \ref{Alg:relay}, $\mid\mathcal{R'}_{k}^{(n)}\mid$ represents the number of elements in set $\mathcal{R'}_{k}^{(n)}$.

\begin{algorithm}[h]
\caption{Find the Optimal Helping Relay Set ${\mathcal{R}_{k}^{(n)}}^*$ for $(k,n)$}
\begin{algorithmic}[1]\label{Alg:relay}
\STATE \textbf{Initialize}:\\
 Set $\mathcal{R}_{k}^{(n)}={1,2,...,K}$;\\
 Remove all relays whose $\gamma_{\textrm{B},r}^{(n)}<\gamma_{\textrm{B},k}^{(n)}$;\\
 Sort the remaining relays in a descending order in terms of $\gamma_{j,k}^{(n)}$, where $j\in \mathcal{R'}_{k}^{(n)}$ and $\gamma_{\textrm{B},j}^{(n)}\geq \gamma_{\textrm{B},k}^{(n)}$;
\FOR{$r=\mid\mathcal{R'}_{k}^{(n)}\mid$ to $1$}
\STATE Calculate $\beta_{\mathcal{R}_{k}^{(n)}}$ associated with $\mathcal{R}_{k}^{(n)}$ in terms of (\ref{Eq:brkn}) and $B_{temp}[\mid\mathcal{R'}_{k}^{(n)}\mid-r+1]=\beta_{\mathcal{R}_{k}^{(n)}}$;\\
\STATE Update $\mathcal{R}_{k}^{(n)}$ by $\mathcal{R}_{k}^{(n)}=\mathcal{R}_{k}^{(n)}-\{r\}$.
\ENDFOR
\STATE Find $\beta_{k,n}^*=\arg\max\limits_{i=1,...,\mid\mathcal{R'}_{k}^{(n)}\mid} B_{temp}[i]$ as the final power gain for user $k$ over $k$ and the corresponding relay set as the optimal helping relay node set ${\mathcal{R}_{k}^{(n)}}^*$ for $(k,n)$ pair.
\end{algorithmic}
\end{algorithm}

By using Algorithm \ref{Alg:relay}, the maximum $\beta_{k,n}^*$ can be determined for each $(k,n)$ pair, which is actually  the maximum power gain for the virtual link associated with the DF relay beamforming between the BS and user $k$ over subcarrier $n$.
For a given $P_{k}^{(n)}$, the maximal achievable information rate for user $k$ over subcarrier $n$ of the DF relay beamforming is
\begin{equation}
{r_{k,n}^{(\textrm{DF})}}^*={\tfrac{W}{2N}\log_2 (1 + 2\beta_{k,n}^*\gamma _{\textrm{B},k}^{(n)}P_{k}^{(n)})}.
\end{equation}
As a result, the maximal achievable information rate over the $n$-th for user $k$ can be given by
 \begin{flalign}\label{Eq:rE2E2}
{r_{k,n}^{(\textrm{E2E})}}^*&= \max
\left\{\tfrac{W}{2N}\log_2 (1 + 2\gamma _{\textrm{B},k}^{(n)}P_{k}^{(n)}),
\tfrac{W}{2N}\log_2 (1 + 2\beta_{k,n}^*\gamma _{\textrm{B},k}^{(n)}P_{k}^{(n)})\right\}\\
&=\tfrac{W}{2N}\log_2 (1 + \max\{\beta_{k,n}^*,1\}2\gamma _{\textrm{B},k}^{(n)}P_{k}^{(n)}).\nonumber
\end{flalign}
Thus, the optimal transmission mode indicator $\rho_{k,n}^*$ can be determined by
 \begin{equation}\label{Eq:rE2E3}
\rho_{k,n}^*= \left\{ {\begin{array}{*{20}{l}}
{0,\,\,\textrm{if}\,\,\beta_{k,n}^*\leq 1}\\
{1,\,\,\textrm{Otherwise}}.
\end{array}} \right.
 \end{equation}

Note that $\rho_{k,n}^*$ also can be determined by the following Corollary 2.

\textbf{Corollary 2.} For a user-subcarrier pair $(k,n)$, if there is no relay $r$ ($1\leq r \leq L$) satisfying that $\gamma_{\textrm{B},r}^{(n)}>\gamma_{\textrm{B},k}^{(n)}$, $\rho_{k,n}^*=0$. Otherwise, $\rho_{k,n}^*=1$.
\begin{proof}
Corollary 2 can be easily proved by using Lemma 1, Lemma 2 and Lemma 4 as follows. If $\gamma_{\textrm{B},r}^{(n)}>\gamma_{\textrm{B},k}^{(n)}$, we can select $r$ as the single helping relay, which results in $\beta_{k,n}^{(n)}$ being larger than 1. In this case, at least $r$ can be added in the helping relay set to achieve a better performance than the DL mode. On the contrary, if no relay satisfies  $\gamma_{\textrm{B},r}^{(n)}>\gamma_{\textrm{B},k}^{(n)}$, according to Lemma 4,  any relay being included in the DF relay beamforming will make it be inferior to the DL mode.
\end{proof}

With the obtained $\rho_{k,n}^*$, we can re-write ${r_{k,n}^{(\textrm{E2E})}}$ as
\begin{equation}\label{Eq:rE2E3}
{r_{k,n}^{(\textrm{E2E})}}=\tfrac{W}{2N}\log_2 (1 + \Upsilon_{\textrm{B},k}^{(n)}P_{k}^{(n)}),
\end{equation}
where
 \begin{equation}\label{Eq:rE2E4}
\Upsilon_{\textrm{B},k}^{(n)}= \left\{ {\begin{array}{*{20}{l}}
{2\gamma _{\textrm{B},k}^{(n)},\quad\,\,\,\,\,\textrm{if}\,\,\rho_{k,n}^*=0}\\
{2\beta_{k,n}^*\gamma _{\textrm{B},k}^{(n)},\,\,\textrm{Otherwise}}.
\end{array}} \right.
 \end{equation}

By doing so, user $k$ can be regarded as being connected with the BS on subcarrier $n$ over an equivalent virtual link with the CNR of $\Upsilon_{\textrm{B},k}^{(n)}$.

\subsubsection{Subcarrier Assignment}

After Step 1, all subcarriers can be assigned in terms of $\Upsilon_{\textrm{B},k}^{(n)}$ over the virtual links.
Actually, to obtain the optimum EE, the subcarrier assignment (i.e., $\theta_{k,n}$) and power consumption (i.e., $ P_{k}^{(n)}$) should be
jointly optimized under the proportional rate constraints.
However, joint optimization of $\theta_{k,n}$ and $P_{k}^{(n)}$ is a combinational problem with very high complexity. Thus, we assign the subcarrier among the users with equal power consumption and allocation over all virtual links at first and then derive the optimal power consumption and allocation over the assigned subcarriers. Such a subcarrier and power separation operation was widely used in the literature to solve the joint subcarrier assignment and power allocation problems for OFDM systems, see e.g., \cite{ProOFDM}. By dosing so, the computational complexity can be greatly reduced.
In this subsection, we present the subcarrier assignment algorithm, and we shall discuss the optimal power consumption and allocation in subsection \ref{Sec:PCA}.

For a given power consumption, to maximize the EE, it is equivalent to maximize the SE of the system. Thus, similar to existing SE maximization-oriented subcarrier assignments, see e.g., \cite{ProOFDM}, our subcarrier assignment also allows each user to select the subchannels with high achievable information rate as much as possible, where the proportional rate fairness among the users is considered. Let $\{\mathcal{S}_k\}$ be the set composed of subcarriers assigned to use $k$. We present the subcarrier assignment as shown in Algorithm \ref{Alg:subcarrier} for clarity.

\begin{algorithm}[h]
\caption{Subcarrier Assignment $\theta_{k,n}^*$ with Equal Power Consumption and Allocation}
\begin{algorithmic}[1]\label{Alg:subcarrier}
\STATE \textbf{Initialize}:\\
 Set ${R_k} = 0$ and $\mathcal{S}_k=\varnothing$, for all $k$ ($k=1,...,K$);\\
 Set $\mathbb{S}={1,...,N}$;\
\FOR{$k=1$ to $K$}
\STATE Find $n=\arg\max\limits_{j\in \mathbb{S}}{r_{k,j}^{(\textrm{E2E})}}$;\\
\STATE Update  ${R_k}=R_k+r_{k,n}^{(\textrm{E2E})}$ according to (\ref{Eq:rE2E3}),  $\mathbb{S} =\mathbb{S} - \{n\}$ and $\mathcal{S}_k=\mathcal{S}_k\cup \{n\}$
\ENDFOR
\WHILE {$\mathbb{S} \ne \emptyset $}
\STATE Find $u=\arg\max\limits_{i=1,2,...,K}{R_i/\alpha_i}$;
\STATE For the found $u$ , $n=\arg\max\limits_{j\in \mathbb{S}}{r_{k,j}^{(\textrm{E2E})}}$;\\
\STATE Update  ${R_k}=R_k+r_{k,n}^{(\textrm{E2E})}$, $\mathbb{S} =\mathbb{S} - \{n\}$ and $\mathcal{S}_k=\mathcal{S}_k\cup \{n\}$
\ENDWHILE
\end{algorithmic}
\end{algorithm}

The subcarrier assignment is suboptimal due to  its utilizing equal power consumption over all subcarriers. Even though, it can averagely achieve the results very close to the optimal solutions, which shall be shown by simulation results in Section V. Note that the subcarrier assignment only provides an rough proportional fairness as it is an integer programming. In the next subsection, we shall present the optimal power consumption over the obtained subcarrier assignment to maximizing the EE of the system while maintaining the proportional rate fairness.

\subsubsection{Optimal Power Consumption and Allocation}\label{Sec:PCA}
In order to achieve higher EE of the system, each user should consume the power resource as less as possible to get higher achievable information rate. By converting the relay links to virtual direct links in Step 1 and with the assigned subcarrier set $\left\{ {\mathcal{S}_k} \right\}$ for all $k=1,2,...,K$ in Step 2, we can transform the optimal problem in (\ref{eq:Opt1}) as
\begin{flalign}\label{eq:Opt2}
\max\limits_{{P_{\textrm{B}}^{(n)}},P_{r,k}^{(n)}}&{\frac{\frac{1}{W}\sum\nolimits_{k=1}^K\sum\limits_{n\in \mathcal{S}_k}\tfrac{W}{2N}\log_2 (1 + \Upsilon_{\textrm{B},k}^{(n)}P_{k}^{(n)})}{\zeta P_{\textrm{trans}}+P_{\textrm{circuit}}}}\\
 \textrm{s.t.}\,\,&(\ref{Con:fairness}), {P_{\textrm{B}}^{(n)}}\geq 0, {P_{r,k}^{(n)}}\geq 0, {P_{\textrm{total}}}\leq P_{\textrm{max}}\nonumber
\end{flalign}
where $\sum\nolimits_{n\in \mathcal{S}_k}\tfrac{W}{2N}\log_2 (1 + \Upsilon_{\textrm{B},k}^{(n)}P_{k}^{(n)})$ is actually the achievable information rate of user $k$, i.e., $R_k$. Since ${R_1}:{R_2}:...:{R_K} = {\alpha _1}:{\alpha _2}:...:{\alpha _K}$, we can introduce the ratio parameter $\delta$ as
\begin{equation}
\delta  = \frac{{{R_1}}}{{{\alpha _1}}} = ... = \frac{{{R_K}}}{{{\alpha _K}}}.
\end{equation}
Then,
Problem (\ref{eq:Opt2}) is further transformed into
\begin{flalign}\label{eq:Opt3}
\max\limits_{{P_{\textrm{B}}^{(n)}},P_{r,k}^{(n)},\delta}\,\,\,&{\frac{\sum\nolimits_{k=1}^K{\delta \alpha_k}}{\zeta P_{\textrm{trans}}+P_{\textrm{circuit}}}}\\
 \textrm{s.t.}\,\,& {P_{\textrm{B}}^{(n)}}\geq 0, {P_{r,k}^{(n)}}\geq 0,{P_{\textrm{total}}}\leq P_{\textrm{max}}\nonumber
\end{flalign}
It can be seen that for a given $\alpha_k$, a larger $\delta$ will lead to a higher $R_k$, requiring more power consumption. This is, all the power elements, ${P_{\textrm{B}}^{(n)}}$, $P_{r,k}^{(n)}$, $P_{k}^{(n)}={P_{\textrm{B}}^{(n)}}+\sum\nolimits_{r\in {\mathscr{R}_{k}^{(n)}}^*\cup \{B\}}P_{r,k}^{(n)}$,
 $P_{\textrm{trans}}$ and $P_{\textrm{circuit}}$, are closely related to $\delta$.
Therefore, they can be described as functions w.r.t $\delta$, i.e., ${P_{\textrm{B}}^{(n)}}(\delta)$, $P_{r,k}^{(n)}(\delta)$, $P_{k}^{(n)}(\delta)$, $P_{\textrm{trans}}(\delta)$ and $P_{\textrm{circuit}}(\delta)$.
As a result, by assuming a quiet large available power $P_{\textrm{max}}$, problem (\ref{eq:Opt3}) can be rewritten as
\begin{flalign}\label{eq:Opt4}
\max\limits_{\delta}&{\frac{\sum\nolimits_{k=1}^K{\delta \alpha_k}}{\zeta P_{\textrm{trans}}(\delta)+P_{\textrm{circuit}}(\delta)}}\\
 \textrm{s.t.}\,\,& \delta> 0.\nonumber
\end{flalign}

In (\ref{eq:Opt4}), the expression of $P_{\textrm{circuit}}(\delta)$ is known, i.e., $P_{\textrm{circuit}}(\delta)= P_{\textrm{static}} + \xi \sum\nolimits_{k=1}^{K}R_{k}=P_{\textrm{static}} + \xi\delta\sum\nolimits_{k=1}^{K}\alpha_{k}$, but the expressions of ${P_{\textrm{B}}^{(n)}}(\delta)$, $P_{r,k}^{(n)}(\delta)$, $P_{k}^{(n)}(\delta)$, $P_{\textrm{trans}}(\delta)$ are unknown.
It can be seen that once the expression of $P_{k}^{(n)}(\delta)$ is derived, ${P_{\textrm{B}}^{(n)}}(\delta)$, $P_{r,k}^{(n)}(\delta)$ can be obtained according to the optimal power allocation in (\ref{Eq:T3}) and the $P_{\textrm{trans}}(\delta)$ can be determined by (\ref{Eq:Pt}).

In the following, we shall deduce the expression of  $P_{k}^{(n)}(\delta)$ and then find the optimal $\delta$ for Problem (\ref{eq:Opt4}).

As is known, given available power $P_{\textrm{trans}}$ for the $K$ users, to achieve the maximum $EE$, the power should be allocated to maximize the total achievable information rate. With the consideration of rate fairness constraint, the power allocation can be formulated as the following optimization problem,
\begin{flalign}\label{eq:Opt5}
\max\limits_{P_{k}^{(n)}}&{\sum\nolimits_{k=1}^K\sum\nolimits_{n\in \mathcal{S}_k}\tfrac{W}{2N}\log_2 (1 + \Upsilon_{\textrm{B},k}^{(n)}P_{k}^{(n)})}\\
 \textrm{s.t.}\,\,&(\ref{Con:fairness}), P_{k}^{(n)}\geq 0,\nonumber\\
 & \sum\nolimits_{k=1}^K\sum\nolimits_{n\in \mathcal{S}_k}P_{k}^{(n)}\leq P_{\textrm{trans}}\nonumber
\end{flalign}
which is a sum-rate maximization problem with proportional rate constraints similar to \cite{ProOFDM}.  The optimization problem in (\ref{eq:Opt5}) is equivalent to find the
maximum of the following Lagrangian function
\begin{flalign}\label{eq:Opt6}
&\mathcal{L}(\mu_1,... \mu_K,P_{k}^{(1)},...,P_{k}^{(\mid \mathcal{S}_k \mid)} )\\
&=\tfrac{W}{2N}\sum\nolimits_{k=1}^K\sum\nolimits_{n\in \mathcal{S}_k}{\log_2 (1 + \Upsilon_{\textrm{B},k}^{(n)}P_{k}^{(n)})}+\mu_1(P_{\textrm{trans}}-\sum\nolimits_{k=1}^K\sum\limits_{n\in \mathcal{S}_k}P_{k}^{(n)})\nonumber\\
&\,\,+\tfrac{W}{2N}\sum\nolimits_{k=2}^K \mu_k\Big(\sum\limits_{n\in \mathcal{S}_1}\log_2(1 + \Upsilon_{\textrm{B},1}^{(n)}P_{k}^{(1)})-\frac{\mu_1}{\mu_k}\sum\limits_{n\in \mathcal{S}_k}\log_2 (1 + \Upsilon_{\textrm{B},k}^{(n)}P_{k}^{(n)})\Big),\nonumber
\end{flalign}
where $\mid \mathcal{S}_k \mid$ denotes the number of elements in set $\mathcal{S}_k$ and $\mu_k$ is the non-negative Lagrangian multiplier. For user $k$, after differentiating it w.r.t. $P_{k}^{(n)}$ and setting the derivatives to be zero, we get
\begin{equation}
\frac{\Upsilon _{\textrm{B},k}^{(m)}}{1 + {P_{k}^{(m)}{\Upsilon _{\textrm{B},k}^{(m)}}}} = \frac{\Upsilon _{\textrm{B},k}^{(n)}}{1 + {P_{k}^{(n)}{\Upsilon _{\textrm{B},k}^{(n)}}}},\,\,\forall \,\,m,n=1,...,\mid\mathcal{S}_k\mid\,\,\textrm{and}\,\,m\neq n.
\end{equation}
Without loss of generality, we  assume that $\Upsilon_{\textrm{B},1}^{(1)}\leq \Upsilon_{\textrm{B},1}^{(2)}\leq \cdot\cdot\cdot \Upsilon_{\textrm{B},1}^{(\mid\mathcal{S}_k\mid)}$.
Then, we can derive the allocated power for  user $k$ over subcarrier $n$ as
\begin{equation}\label{Eq:pkn3}
{P_{k}^{(n)}} = {P_{k}^{(1)}} + \frac{\Upsilon _{\textrm{B},k}^{(n)} - \Upsilon _{\textrm{B},k}^{(1)}}{\Upsilon _{\textrm{B},k}^{(n)}\Upsilon _{\textrm{B},k}^{(1)}},\,\,\forall \,\,n=2,...,\mid\mathcal{S}_k\mid.
\end{equation}
According to (\ref{Eq:rE2E3}), $P_{k}^{(n)}$ can be expressed
by
\begin{equation}\label{Eq:pkn4}
P_{k}^{(n)}=\frac{2^{\frac{2N}{W}{r_{k,n}^{(\textrm{E2E})}}}-1}{\Upsilon_{\textrm{B},k}^{(n)}},\,\,\forall \,\,n=2,...,\mid\mathcal{S}_k\mid
\end{equation}
Combining (\ref{Eq:pkn3}) and (\ref{Eq:pkn4}), we obtain that
\begin{equation}\label{Eq:pkn5}
{P_{k}^{(n)}} = \frac{2^{\frac{2N}{W}{r_{k,1}^{(\textrm{E2E})}}}-1}{\Upsilon_{\textrm{B},k}^{(1)}}+ \frac{\Upsilon _{\textrm{B},k}^{(n)} - \Upsilon _{\textrm{B},k}^{(1)}}{\Upsilon _{\textrm{B},k}^{(n)}\Upsilon _{\textrm{B},k}^{(1)}},\,\,\forall \,\,n=2,...,\mid\mathcal{S}_k\mid
\end{equation}
Therefore, for all $n=2,...,\mid\mathcal{S}_k\mid$, we have that \begin{equation}\label{Eq:rr1}
\frac{2^{\frac{2N}{W}{r_{k,n}^{(\textrm{E2E})}}}-1}{\Upsilon_{\textrm{B},k}^{(n)}}= \frac{2^{\frac{2N}{W}{r_{k,1}^{(\textrm{E2E})}}}-1}{\Upsilon_{\textrm{B},k}^{(1)}}+ \frac{\Upsilon _{\textrm{B},k}^{(n)} - \Upsilon _{\textrm{B},k}^{(1)}}{\Upsilon _{\textrm{B},k}^{(n)}\Upsilon _{\textrm{B},k}^{(1)}}.
\end{equation}

\textbf{Lemma 5.} \textit{For a given $\delta$, the optimally allocated power associated with user $k$ over subcarrier $n$ is}
\begin{flalign}\label{Eq:pknd}
{P_{k}^{(n)}} = \frac{2^{\frac{2N}{W}{\frac{1}{\mid\mathcal{S}_k\mid}(\delta\alpha_K-Q_k)}}-1}{\Upsilon_{\textrm{B},k}^{(1)}}+ \frac{\Upsilon _{\textrm{B},k}^{(n)} - \Upsilon _{\textrm{B},k}^{(1)}}{\Upsilon _{\textrm{B},k}^{(n)}\Upsilon _{\textrm{B},k}^{(1)}},
\end{flalign}
\textit{and the total power allocated to user $k$
is}
\begin{flalign}\label{Eq:pktot}
{P_{k,\textrm{tot}}} = \frac{\mid\mathcal{S}_k\mid2^{\frac{2N}{W}{\frac{1}{\mid\mathcal{S}_k\mid}(\delta\alpha_K-Q_k)}}-\mid\mathcal{S}_k\mid}{\Upsilon_{\textrm{B},k}^{(1)}}+ \sum\limits_{n\in \mathcal{S}_k}\frac{\Upsilon _{\textrm{B},k}^{(n)} - \Upsilon _{\textrm{B},k}^{(1)}}{\Upsilon _{\textrm{B},k}^{(n)}\Upsilon _{\textrm{B},k}^{(1)}},
\end{flalign}
\textit{where} $Q_k\triangleq\sum\limits_{n \in \mathcal{S}_k}\tfrac{W}{2N}\log_2\Bigg( \frac{\Upsilon_{\textrm{B},k}^{(n)}}{\Upsilon_{\textrm{B},k}^{(1)}}\Bigg)$
\begin{proof}
From (\ref{Eq:rr1}), it is deduced that
\begin{equation}\label{Eq:rr2}
\frac{2^{\frac{2N}{W}{r_{k,n}^{(\textrm{E2E})}}}}{\Upsilon_{\textrm{B},k}^{(n)}}= \frac{2^{\frac{2N}{W}{r_{k,1}^{(\textrm{E2E})}}}}{\Upsilon_{\textrm{B},k}^{(1)}},
\end{equation}
which means that
\begin{equation}\label{Eq:rr3}
r_{k,n}^{(\textrm{E2E})}=\tfrac{W}{2N}\log_2\Bigg( \tfrac{\Upsilon_{\textrm{B},k}^{(n)}}{\Upsilon_{\textrm{B},k}^{(1)}}\Bigg)+r_{k,1}^{(\textrm{E2E})}.
\end{equation}
Thus,
\begin{flalign}\label{Eq:Rkk}
R_k&=\delta \alpha_K=\sum\limits_{n \in \mathcal{S}_k} r_{k,n}^{(\textrm{E2E})}=\mid\mathcal{S}_k\mid r_{k,1}^{(\textrm{E2E})}+\sum\limits_{n \in \mathcal{S}_k-\{1\}}\tfrac{W}{2N}\log_2\Bigg( \tfrac{\Upsilon_{\textrm{B},k}^{(n)}}{\Upsilon_{\textrm{B},k}^{(1)}}\Bigg).
\end{flalign}
As a result,
\begin{flalign}\label{Eq:Rkk2}
r_{k,1}^{(\textrm{E2E})}=\frac{1}{\mid\mathcal{S}_k\mid}\left(\delta\alpha_K-\sum\limits_{n \in \mathcal{S}_k-\{1\}}\tfrac{W}{2N}\log_2\left( \tfrac{\Upsilon_{\textrm{B},k}^{(n)}}{\Upsilon_{\textrm{B},k}^{(1)}}\right)\right)=\frac{1}{\mid\mathcal{S}_k\mid}(\delta\alpha_K-Q_k)
\end{flalign}
Combining (\ref{Eq:Rkk2}) with (\ref{Eq:pkn5}), we can obtain (\ref{Eq:pknd}). Moreover, as
\begin{flalign}\label{Eq:Rtot1}
{P_{k,\textrm{tot}}}=\sum\nolimits_{n \in \mathcal{S}_k} P_{k}^{(n)},
\end{flalign}
we can arrive at (\ref{Eq:pktot}) by substituting (\ref{Eq:pknd}) into (\ref{Eq:Rtot1}). Therefore, Lemma 5 is proved.
\end{proof}

Based on Lemma 5, we obtain the Corollary 3 as follows.

\textbf{Corollary 3.}\textit{ For a given $\delta$, the total transmit power of the relay aided OFDM system can be expressed by}
\begin{flalign}\label{Eq:Ptransd}
P_{\textrm{trans}}(\delta)&=\sum\nolimits_{k=1}^K {P_{k,\textrm{tot}}}\\
&= \sum\nolimits_{k=1}^K\left (\frac{\mid\mathcal{S}_k\mid2^{\frac{2N}{W}{\frac{1}{\mid\mathcal{S}_k\mid}\left(\delta\alpha_K-\sum\limits_{n \in \mathcal{S}_k}\tfrac{W}{2N}\log_2\left( \frac{\Upsilon_{\textrm{B},k}^{(n)}}{\Upsilon_{\textrm{B},k}^{(1)}}\right)\right)}}-\mid\mathcal{S}_k\mid}{\Upsilon_{\textrm{B},k}^{(1)}}+ \sum\limits_{n\in \mid\mathcal{S}_k\mid}\frac{\Upsilon _{\textrm{B},k}^{(n)} - \Upsilon _{\textrm{B},k}^{(1)}}{\Upsilon _{\textrm{B},k}^{(n)}\Upsilon _{\textrm{B},k}^{(1)}}\right).\nonumber
\end{flalign}
\begin{proof}
Corollary 3 can be proved by combining (\ref{Eq:pktot}) with $P_{\textrm{trans}}(\delta)=\sum\nolimits_{k=1}^K {P_{k,\textrm{tot}}}$.
\end{proof}

With Corollary 3, we express $EE$ as a function of $\delta$ as
\begin{equation}
EE\left( \delta  \right) = \frac{\sum\nolimits_{k=1}^K{\delta \alpha_k}}{\zeta P_{\textrm{trans}}(\delta)+P_{\textrm{static}} + \xi \sum\nolimits_{k=1}^{K}\delta \alpha_k}.
\end{equation}

\textbf{Theorem 4.} \textit{$EE\left( \delta  \right)$ is a first increasing and then decreasing function w.r.t. $\delta$ and there exists an optimal
$\delta^\sharp \in [0,+\infty)$ such that $EE\left( \delta  \right)$ achieves the maximum.}

\begin{proof}
The derivative of $EE$  w.r.t $\delta $  is given by
\begin{flalign}\label{eq:12}
\frac{{\textrm{d}EE\left( \delta  \right)}}{{\textrm{d}\delta }}&= \frac{\sum\limits_{k = 1}^K {{\alpha _k}} \left( {\zeta P_{\textrm{trans}}\left( \delta  \right) + {P_{\textrm{static}} + \xi \sum\limits_{k=1}^{K}\delta \alpha_k}} \right) - \left( {\delta \sum\limits_{k = 1}^K {{\alpha _k}} } \right)\left(\zeta \frac{\textrm{d} P_{\textrm{trans}}(\delta)}{\textrm{d} \delta}+\xi \sum\limits_{k=1}^{K} \alpha_k\right)}{{{{\left( {\zeta P_{\textrm{trans}}\left( \delta  \right) + {P_{\textrm{circuit}}}} \right)}^2}}}\nonumber\\
&= \frac{{\sum\nolimits_{k = 1}^K {{\alpha _k}} }\left( {\zeta P_{\textrm{trans}}\left( \delta  \right) + {P_{\textrm{static}} + \xi \sum\nolimits_{k=1}^{K}\delta \alpha_k} - \delta \zeta \frac{\textrm{d} P_{\textrm{trans}}(\delta)}{\textrm{d} \delta}}-\delta\xi \sum\nolimits_{k=1}^{K}\alpha_k \right)}{{{{\left( {\zeta P_{\textrm{trans}}\left( \delta  \right) + {P_{\textrm{static}} + \xi \sum\nolimits_{k=1}^{K}\delta \alpha_k}} \right)}^2}}}\nonumber\\
&= \frac{{\sum\nolimits_{k = 1}^K {{\alpha _k}} }}{{{{\left( {\zeta P_{\textrm{trans}}\left( \delta  \right) + {P_{\textrm{static}} + \xi \sum\nolimits_{k=1}^{K}\delta \alpha_k}} \right)}^2}}}\left( {\zeta P_{\textrm{trans}}\left( \delta  \right) + {P_{\textrm{static}} } - \delta \zeta \frac{\textrm{d} P_{\textrm{trans}}(\delta)}{\textrm{d} \delta}} \right)
\end{flalign}
It is observed that the sign of $\frac{\textrm{d}EE( \delta)}{\textrm{d}\delta }$ is determined by the term $\zeta P_{\textrm{trans}}\left( \delta  \right) + {P_{\textrm{circuit}}} - \delta \zeta
\frac{\textrm{d} P_{\textrm{trans}}(\delta)}{\textrm{d} \delta}$. Let $G(\delta)=\zeta P_{\textrm{trans}}\left( \delta  \right) + {P_{\textrm{circuit}}} - \delta \zeta \frac{\textrm{d} P_{\textrm{trans}}(\delta)}{\textrm{d} \delta}$. We have that $\frac{d G(\delta)}{d \delta}=-\delta \zeta \tfrac{\textrm{d} P_{\textrm{trans}}^2(\delta)}{\textrm{d} \delta^2}$. Since
\begin{flalign}\label{Eq:Ptransd}
\frac{\textrm{d} P_{\textrm{trans}}^2(\delta)}{\textrm{d} \delta^2}= \sum\limits_{k=1}^K
\frac{1}{\mid\mathcal{S}_k\mid\Upsilon _{\textrm{B},k}^{(1)}2^{\frac{\frac{2N}{W}Q_k-\alpha_k\frac{2N}{W}\delta }{\mid\mathcal{S}_k\mid}\alpha_k^2(\frac{2N}{W})^2(\ln2)^2}},
\end{flalign}
which is always larger than 0, it can be inferred that $\frac{d G(\delta)}{d \delta}<0$. It means that $G(\delta)$ is a decreasing function w.r.t. $\delta$ over $0\leq \delta \leq +\infty$.
Moreover, $G(\delta=0)>0$. Therefore, there must exist a $\delta^\sharp$ such that when $ \delta = \delta^\sharp$, $G(\delta)=0$, when $0\leq \delta < \delta^\sharp$, $G(\delta)>0$ and when $\delta^\sharp < \delta$, $G(\delta)<0$. As a result, it can be inferred that when $ \delta = \delta^\sharp$, $\frac{{dEE\left( \delta  \right)}}{{d\delta }}=0$, when $0\leq \delta < \delta^\sharp$, $\frac{{d EE\left( \delta  \right)}}{{d\delta }}>0$ and when $\delta^\sharp < \delta$, $\frac{{dEE\left( \delta  \right)}}{{d\delta }}<0$.
Thus, $\delta^\sharp$ is just the optimal $\delta^\sharp$ which makes $ EE\left( \delta  \right)$ achieve the maximum.
\end{proof}

Note that although there is an optimal $\delta^\sharp$  achieving the maximum $EE$, $\delta^\sharp$ is obtained without considering the constraint $P_{\textrm{max}}$ and
$P_{\textrm{total}}(\delta^\sharp)=P_{\textrm{trans}}(\delta^\sharp)+P_{\textrm{circuit}}(\delta^\sharp)$ may not always satisfy the given available power constraint $P_{\textrm{max}}$. According to Theorem 4 that $EE\left( \delta  \right)$ is a first increasing and then decreasing function and when $\delta=\delta^\sharp$ it achieves the maximum, it can be deduced that if $P_{\textrm{total}}(\delta^\sharp)<P_{\textrm{max}}$, $EE\left( \delta  \right)$ achieves the maximum at $P_{\textrm{total}}(\delta^\sharp)$, which means that in this case when system achieve the maximum EE, the available power, i.e., $P_{\textrm{max}}$, is not run out. Otherwise, $EE\left( \delta  \right)$ achieves the maximum at $P_{\textrm{max}}$, which means that in this case, to achieve the maximum EE all available power $P_{\textrm{trans}}$ is run out. Based on above analysis, the optimal $\delta^*$ with the available power constraint $P_{\textrm{trans}}\leq P_{\textrm{max}}$  can be determined as follows.
\begin{equation}\label{Eq:optdelta}
\delta^*= \left\{ {\begin{array}{*{20}{l}}
{\delta^\sharp ,\,\,\textrm{if}\,\,P_{\textrm{trans}}(\delta^\sharp )<P_{\textrm{max}}}\\
{\delta^\diamond,\,\,\textrm{Otherwise}}.
\end{array}} \right.
 \end{equation}
where $\delta^\sharp $ and $\delta^\diamond$ can be obtained by solving the equation $\zeta P_{\textrm{trans}}\left( \delta^\sharp   \right) + {P_{\textrm{circuit}}} - \delta^\sharp  \zeta \frac{\textrm{d} P_{\textrm{trans}}(\delta)}{\textrm{d} \delta} \left( \delta^\sharp   \right)=0$ and  $\zeta P_{\textrm{trans}}\left( \delta^\diamond  \right) + {P_{\textrm{circuit}}} - \delta^\diamond \zeta \frac{\textrm{d} P_{\textrm{trans}}(\delta)}{\textrm{d} \delta}\left( \delta^\diamond  \right)=P_{\textrm{max}}$, respectively, by using some efficient root-finding methods such as Bisection algorithm and Newton's algorithm, etc.

As shown in Figure \ref{Fig:fowchart}, the initial subcarrier assignment is obtained by assuming that all available power  $P_{\textrm{max}}$ is used up, i.e.,  $P_{\textrm{trans}}=\frac{P_{\textrm{max}}-P_{\textrm{static}}}{\zeta}$. According to the analysis in this subsection, the actual power consumed to obtain maximum EE may not be the same with the constrained available power $P_{\textrm{max}}$. Sometimes, it is less than $P_{\textrm{max}}$, i.e., $P_{\textrm{trans}}(\delta^\sharp )<P_{\textrm{max}}$. Considering that the assumption of $P_{\textrm{trans}}=\frac{P_{\textrm{max}}-P_{\textrm{static}}}{\zeta}$ may result in an inaccurate subcarrier assignment, as shown in Figure \ref{Fig:fowchart}, we update  $P_{\textrm{trans}}$ with $P_{\textrm{trans}}(\delta^\sharp )$ and re-execute step 2 and step 3 circularly in our proposed resource allocation. By introducing such an interaction operation of step 2 and step 3, a better EE performance is expected to be achieved. However, with simulations, we found that in most cases, step 2 and step 3 are executed only once and in the rest minority cases, step 2 and step 3 are required to be executed not more than twice.

\subsection{Computational Complexity Analysis}\label{Sec:CA}

In this part, we shall analyze the computational complexity of our proposed resource allocation. For comparison, the intuitive method with computer search is also analyzed.

In our proposed resource allocation, Step 1 calculates the optimal ${\mathcal{R}_{k}^{(n)}}^*$ for all $(k, n)$ pairs. The computational complexity is $\mathcal {O}(KNL)$. Step 2 gets the optimal subcarrier assignment for a given power consumption. The computational complexity is $\mathcal {O}(KN)$. Step 3 calculates the optimal power consumption and allocation over the assigned subcarriers, where some explicit results are used. Thus, the computational complexity of Step 3 is only determined by the calculation of $\delta^*$ whose complexity is only determined by the root-finding method and has no relation ship with $K$, $L$ and $N$. So, the computational complexity associated with Step 3 is $\mathcal {O}(1)$. Besides, in our proposed resource allocation scheme, Step 2 and Step 3 are executed circularly as shown in Figure \ref{Fig:fowchart}. Therefore, the total complexity of our proposed resource allocation is $\mathcal {O}(\kappa(KNL+KN))$. $\kappa$ denote the number of times of the circling execution of Step 2 and Step 3.

Comparatively, if the intuitive method is used, although the optimal result can be guaranteed, it is too complex for practical use, especially for relatively large $N$ and $K$ scenarios. In the intuitive method, for each $(k,n)$, all possible relay¡¡sets, $2^L$ relay sets, have to be checked, which is with the computational complexity of  $\mathcal {O}(2^L)$. Thus, finding ${\mathcal{R}_{k}^{(n)}}^*$ for all $(k,n)$ pairs results in a computational complexity of  $\mathcal {O}(KN2^L)$. Moreover, the power consumption is also performed over all possible subcarrier assignments, i.e., $K^N$ possible subchannel assignments. Therefore, the computational complexity of the intuitive method is at least $\mathcal {O}(KN2^L+K^N)$. That is, our proposed scheme is at least in the order of $K^N$ times less than that of the optimal one.


\section{Simulation Results and Discussions}
In this section, we shall provide some simulation results to evaluate the performance of our proposed resource allocation scheme. We shall also discuss the EE behaviors of the system via numerical results and then present some useful insights.

In the simulations, the wireless channel is modelled as a frequency-selective channel so that the signal delivered on each subcarrier undergoes identical Rayleigh fading independently. The total available bandwidth and transmit power are 1 MHz and 1 Watt, respectively. ${P_{\textrm{static}}}=0.2$ Watt and $\eta=0.38$. These configurations will not change throughout the following simulations unless otherwise specified.

\begin{figure}
\centering
\includegraphics[width=0.47\textwidth]{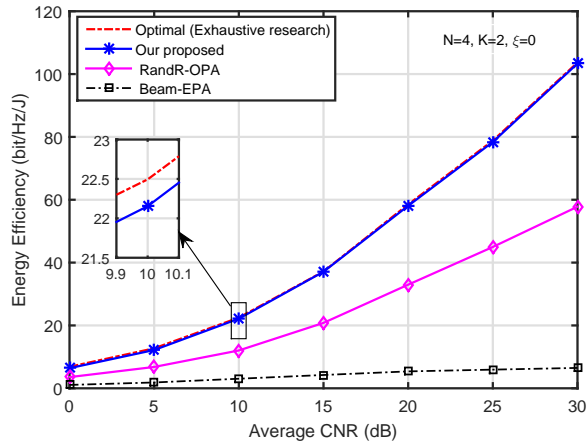}
\caption{EE performance comparison with the optimal results for $\xi=0$ case.}
\label{Fig:ComOptx0}
\end{figure}

\begin{figure}
\centering
\includegraphics[width=0.47\textwidth]{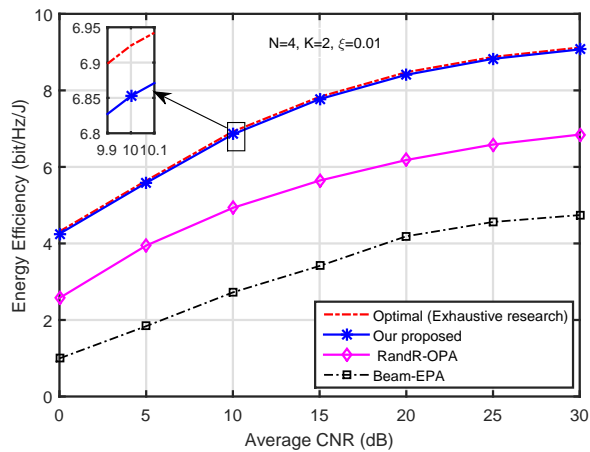}
\caption{EE performance comparison with the optimal results for $\xi=0.01$ case.}
\label{Fig:ComOptxR}
\end{figure}

Figure \ref{Fig:ComOptx0} and Figure \ref{Fig:ComOptxR} compare the maximal EE achieved by our proposed resource allocation with the optimal ones obtained by computer search for $\xi=0$ and $\xi=0.01$, respectively. In both figures, $N=4$, $L=5$ and $K=2$. $\alpha_1:\alpha_2=1:1$. Each plotted point on the curves was averaged over 1000 simulations. For comparison, we also plot the EE performance of two benchmark methods, i.e., \emph{RandR-OPA} and \emph{Beam-EPA}, where in RandR-OPA, only one relay is randomly selected and the optimal power allocation is adopted, and in Beam-EPA, the DF relay beamforming is used but with equal power allocation among all nodes. From the two figures, it can be observed that the maximum EE obtained by our proposed scheme is very close to the optimal ones obtained by the exhaustive computer search, which demonstrates that by adopting our proposed scheme, the approximately optimum EE performance can be achieved for the multi-relay OFDM system. Moreover, it also can be seen that with the increment of average CNR, the EE performance of the system is increased, where the EE reaches a relatively high level in the high CNR regime. This is because to achieve the same information rate, in high CNR regime, the power required is less than that in low CNR regime, which yields higher $\delta$ and EE in relatively high CNR regime.

Figure \ref{Fig:ComOptx0} and Figure \ref{Fig:ComOptxR} also show that different circuit power consumption leads to very different system EE performance behavior. In Figure \ref{Fig:ComOptx0}, $\xi=0$, which means that the circuit power consumption is treated as constant. In this case, the growth rate of the system maximum EE increases with the increment of average CNR. Relatively, in Figure \ref{Fig:ComOptx0}, $\xi=0.01$, which means that the circuit power consumption linearly increase of the total information rate. In this case, the growth rate of the system maximum EE decreases with the increment of average CNR.

\begin{figure}
\centering
\includegraphics[width=0.47\textwidth]{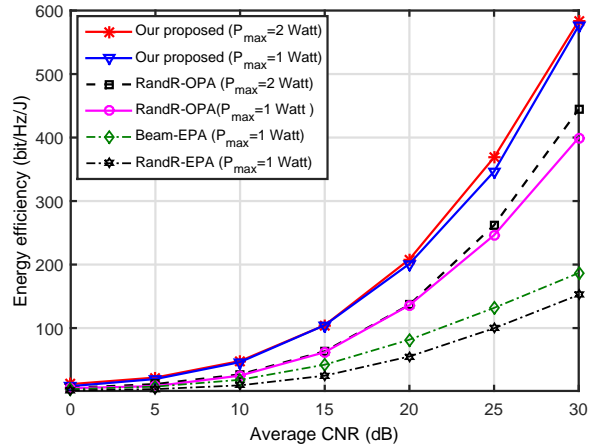}
\caption{EE performance comparisons versus average CNR with $\xi=0$, where $N=50$, $L=20$ and $K=10$.}
\label{Fig:Comx0}
\end{figure}

\begin{figure}
\centering
\includegraphics[width=0.47\textwidth]{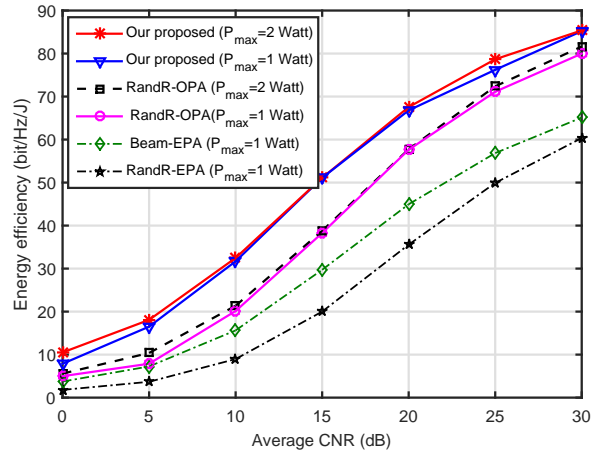}
\caption{EE comparisons versus average CNR with $\xi=0.01$, where $N=50$, $L=20$ and $K=10$.}
\label{Fig:ComxR}
\end{figure}

In order to perform more sufficient comparisons, in Figure \ref{Fig:Comx0} and Figure \ref{Fig:ComxR}, we compare our proposed scheme with the benchmark shemes, i.e., RandR-OPA and Beam-EPA, for a system configuration with more subcarriers, more relays and more users. In Figure \ref{Fig:Comx0} and Figure \ref{Fig:ComxR}, $\xi$ is set to be 0 and 0.01, respectively. In both figures, $N$, $L$ and $K$ are increased to be 50, 20 and 10, respectively. The proportional rate fairness constraints among the 10 users are set to be $\alpha_1: \alpha_2: \alpha_3: \alpha_4: \alpha_5: \alpha_6: \alpha_7: \alpha_8: \alpha_9: \alpha_{10}= 1: 1 : 2: 3: 4 :5: 5: 6: 6: 7$. From Figure \ref{Fig:Comx0} and Figure \ref{Fig:ComxR}, we can seen that our proposed scheme always achieves EE performance gain compared with RandR-OPA and Beam-EPA. Similar to Figure \ref{Fig:ComOptx0} and Figure \ref{Fig:ComOptxR}, the maximum EE performance behavior of the system is greatly affected by the circuit power consumption. That is, for the constant circuit power consumption, system EE increasing rate is an increasing function of the system average CNR, while for the linearly rate-dependent one, system EE increasing rate is a decreasing function of the system average CNR.
\begin{figure}
\centering
\includegraphics[width=0.47\textwidth]{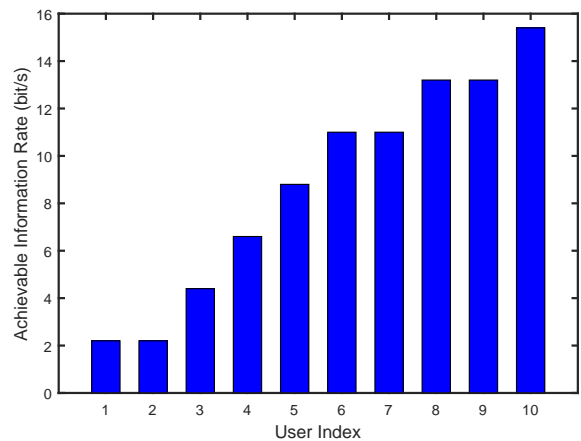}
\caption{EE comparison with the optimal results with $\xi=0$.}
\label{Fig:UserFairness}
\end{figure}
Figure \ref{Fig:UserFairness} plots the achievable information obtained my our scheme for the 10 users corresponding to Figure \ref{Fig:Comx0}. It can be seen that with our proposed scheme, the proportional rate fairness among the users can be guaranteed.

\begin{figure}
\centering
\includegraphics[width=0.47\textwidth]{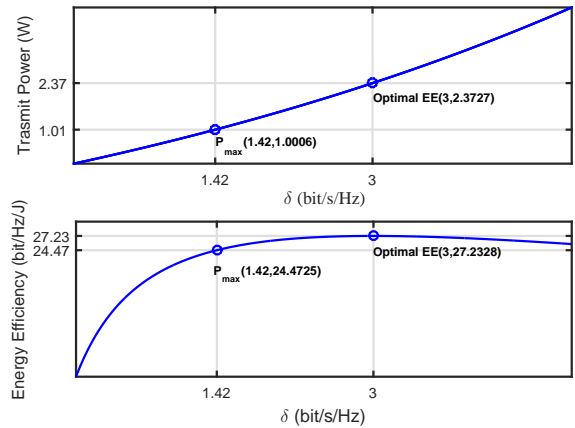}
\caption{SE and EE verus $\delta$.}
\label{Fig:Delta}
\end{figure}
\begin{figure}
\centering
\includegraphics[width=0.47\textwidth]{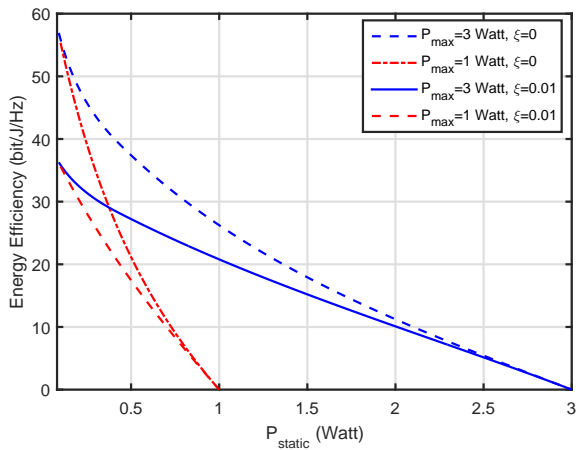}
\caption{EE performance versus $P_{\textrm{static}}$.}
\label{Fig:Pstatic}
\end{figure}
\begin{figure}
\centering
\includegraphics[width=0.47\textwidth]{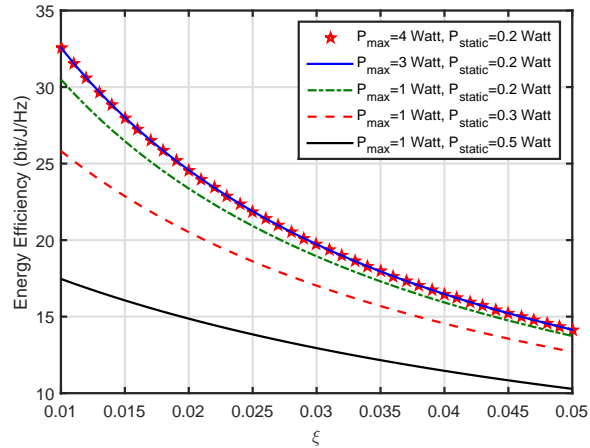}
\caption{EE performance versus $\xi$.}
\label{Fig:xi}
\end{figure}
\begin{figure}
\centering
\includegraphics[width=0.47\textwidth]{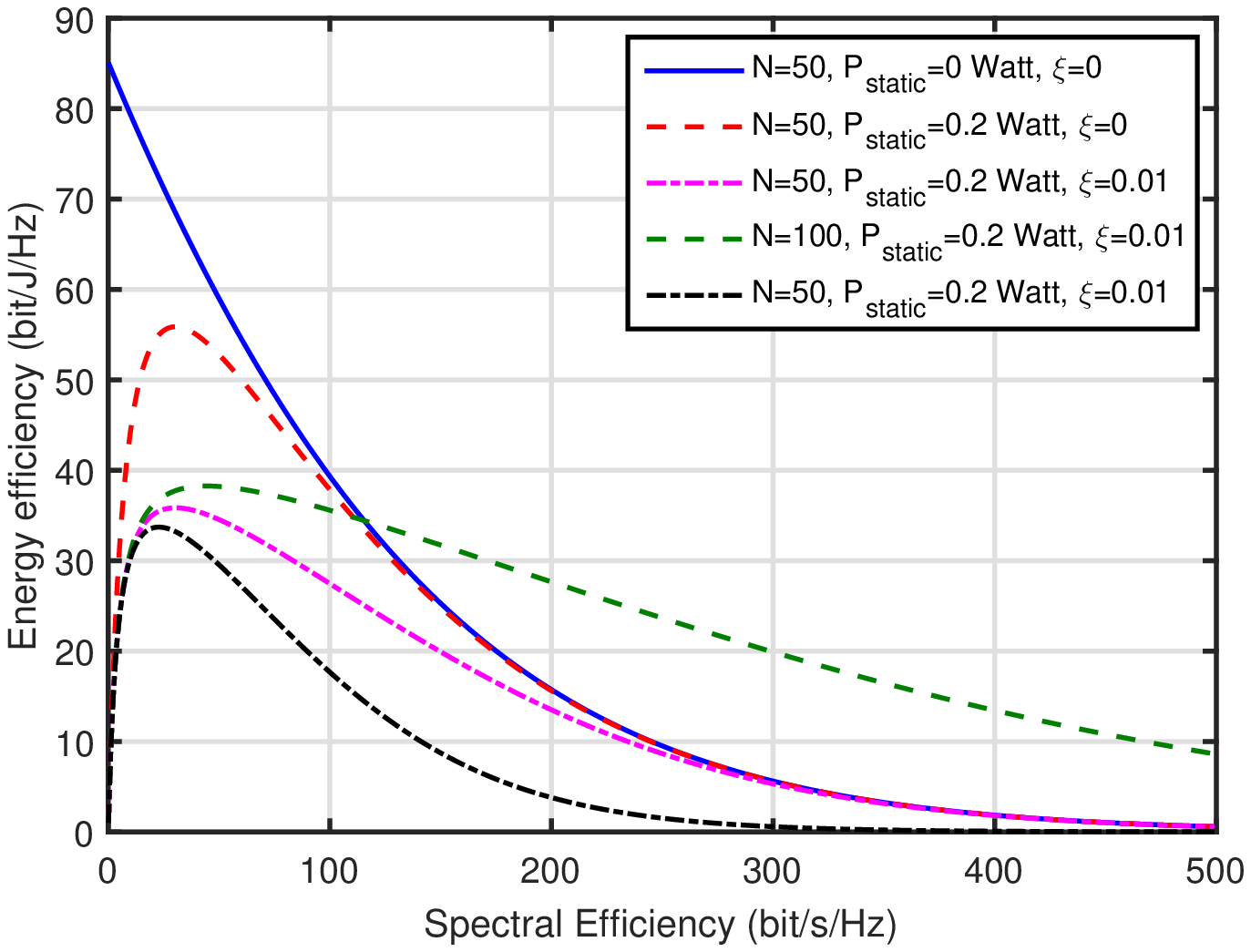}
\caption{The trade-off between the EE and SE of the system.}
\label{Fig:Trade_OFF}
\end{figure}
\begin{figure}
\centering
\includegraphics[width=0.47\textwidth]{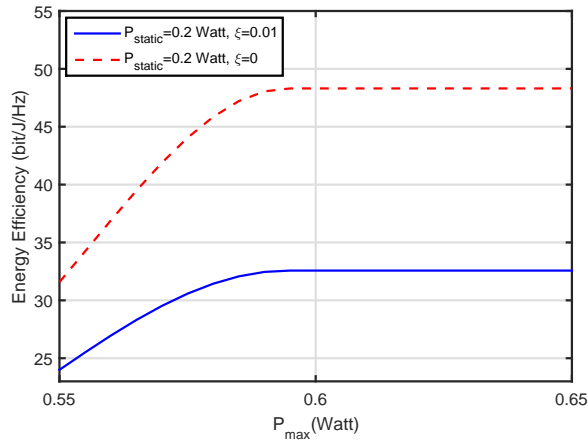}
\caption{EE performance versus $P_{\textrm{max}}$.}
\label{Fig:Pmax}
\end{figure}

To better understand the relationship between EE and the ratio parameter $\delta$, we plot SE and EE verus $\delta$ in Figure \ref{Fig:Delta}, where we set $N$ and $K$ to be 50 and 5, respectively. The average CNR is 10dB. The proportional rate constraints are set to 1:1:1:1:1, which indicates that each user has the same information rate service priority. Figure \ref{Fig:Delta} shows that the transmit power $P(\delta)$ is monotonously increasing in $\delta$ and the $EE(\delta)$ is first strictly increasing and then strictly decreasing w.r.t $\delta$. This result is consistent with our obtained theoretical result in Theorem 4. In Figure \ref{Fig:Delta}, it also can be observed that when $P_{\textrm{max}}=$ 1 Watt,  the optimal $\delta$ equals 1.42 bit/s/Hz. In this case, all power is used up, leading to the maximal achievable EE as 24.4725 bit/Hz/J. If we set $P_{\textrm{max}}$ to 3 Watt, which is larger than the EE maximized transmit consumption, i.e., 2.36 Watt, the maximum EE of 27.2328 bit/Hz/J can be achieved. This observation also proves  the
correctness of the analysis in section \ref{Sec:PCA}.
Figure \ref{Fig:Pstatic} and Figure \ref{Fig:xi} present the EE performance of the system versus $P_{\textrm{static}}$ and $\xi$, respectively, where $P_{\textrm{max}}$ is set to $1$ Watt. It shows that both the increment $P_{\textrm{static}}$ and $\xi$ will lead to the decrease of the system EE and the decreasing rate is decrease with the growth of $P_{\textrm{static}}$ and $\xi$. The reason is that higher $P_{\textrm{static}}$ and $\xi$ represent higher circuit power consumption. For given $P_{\textrm{max}}$, the power  used for information transmission is decreased so that the achievable information rate is also decreased. Figure \ref{Fig:Trade_OFF} shows the trade-off relationship between EE and SE of the system. For the ideal circuit case, i.e., $P_{\textrm{static}}=0$ and $\xi=0$ case, EE strictly decreases with the increase of SE. The reason is that higher SE will consume more power which may lead to the decrease of EE. For the non ideal circuit cases, wether $P_{\textrm{circuit}}=P_{\textrm{static}}$ as a constant or $P_{\textrm{circuit}}=P_{\textrm{static}}+\xi\sum_{k=1}^{K}R_k$ as a linear function of $\xi$, EE first increases and then decreases with the increment of SE.
Figure \ref{Fig:Pmax} presents the EE performance of the system versus $P_{\textrm{max}}$. It can be seen that the maximum EE of the system first increases and then keeps stable with the increment of $P_{\textrm{max}}$. The reason is that when $P_{\textrm{max}}$ is relatively small, all $P_{\textrm{max}}$ is consumed to achieve the maximum EE for the system. in this case, the increment of $P_{\textrm{max}}$ can increase the system EE. But for a relatively large $P_{\textrm{max}}$, similar to the results in Figure
\ref{Fig:Delta}, when $P_{\textrm{max}}>P_{\textrm{total}}(\delta^*)$, although $P_{\textrm{max}}$ is increased, i.e., $\delta$ is increased, the maximum EE can not be raised.

\begin{figure}
\centering
\includegraphics[width=0.47\textwidth]{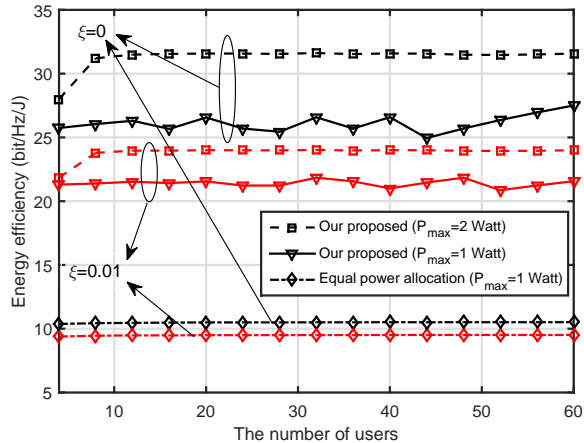}
\caption{EE performance versus $K$.}
\label{Fig:User}
\end{figure}

\begin{figure}
\centering
\includegraphics[width=0.47\textwidth]{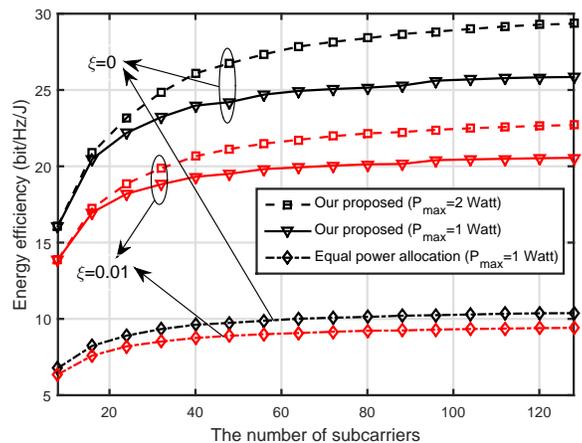}\picspace
\caption{EE performance versus $N$.}
\label{Fig:subcarrier}
\end{figure}

Figure \ref{Fig:User} and Figiure \ref{Fig:subcarrier} plot the EE performance versus $K$ and $N$, respectively. Figure \ref{Fig:User}  shows that the EE of the system is not affected by the number of user notably. The reason is that $K$ has little effects on the power allocation, which may not impact the system performance obviously. Figure \ref{Fig:subcarrier} shows that with the increment of $N$, the EE performance can be increased. But the increment rate is decrease with the growth of $N$. The reason is that  more subcarriers bring more subchannel diversity and more subcarrier assignment flexibility which may increase the system EE performance but more subcarriers result in less bandwidth allocated to each subcarrier which may decrease the improving gain of the EE performance.

\section{Conclusion}

This paper studied the EE performance in multi-relay OFDM systems with DF relay beamforming employed. In order to explore the EE performance with user fairness for such a system, we formulated an optimization problem to maximize the EE by jointly optimizing the transmission mode selection, the helping relay set selection, the subcarrier assignment and the power consumption and allocations under nonlinear proportional rate constraints, where both the transmit power consumption and the linearly rate-dependent circuit power consumption were considered. As the problem cannot be directly solved, we presented an low-complexity scheme for it. Simulation results demonstrate that our proposed resource allocation can achieve the approximating optimum. It is also shown that with both the constant and the linearly rate-dependent circuit power consumption the system EE grows with the increment of system average CNR, but the growth rates are very different. For the constant circuit power consumption, system EE growth rate is a increasing function of CNR while for the linearly rate-dependent one, system EE growth rate is a decreasing function of CNR.  This observation is insightful which indicates that by deducing the circuit dynamic power consumption per unit data rate, system EE can be greatly enhanced. Besides, we also discussed the effects of the number of users and subcarriers on the system EE performance.



%

\end{document}